\documentclass[preprint, review, 12pt]{elsarticle}
\usepackage{amssymb}
\usepackage{amsmath} 
\usepackage[ruled]{algorithm2e}
\usepackage{colortbl,booktabs} 
\usepackage{tabularx}
\usepackage{threeparttable}
\usepackage{multicol,multirow}
\usepackage{graphicx}
\usepackage{subfigure} 
\usepackage{enumerate}
\usepackage{float}
\usepackage{graphicx}
\usepackage{caption}
\usepackage [font=footnotesize,labelfont=bf]{caption}
\usepackage{rotating}
\usepackage{tikz}
\usepackage[colorlinks,linkcolor=blue]{hyperref}
\usepackage{color}
\definecolor{R}{rgb}{1, 0.0, 0.0}
\definecolor{G}{rgb}{0.0, 1, 0.0}
\definecolor{B}{rgb}{0.0, 0.0, 1}
\journal{Physics of Fluids}

\begin{document}
\begin{frontmatter}
	\title{Corrected Riemann smoothed particle hydrodynamics 
		method for multi-resolution fluid-structure interaction}
	\author[author1]{Bo Zhang }
	\ead{bo.zhang.aer@tum.de}
	\author[author2]{Jianfeng Zhu }
	\author[author1]{Xiangyu Hu \corref{mycorrespondingauthor}}
	\cortext[mycorrespondingauthor]{Corresponding author.}
	\ead{xiangyu.hu@tum.de}
	\address[author1]{TUM School of Engineering and Design, 
		Technical University of Munich,\\
		Garching 85748, Germany}
	\address[author2]{School of Aerospace Engineering, 
		Xiamen University,\\
		Xiamen 361005, China}
	\begin{abstract}
		As a mesh-free method, smoothed particle hydrodynamics 
		(SPH) has been widely used for modeling and simulating 
		fluid-structure interaction (FSI) problems.
		While the kernel gradient correction (KGC) method is 
		commonly applied in structural domains to enhance 
		numerical consistency, high-order consistency 
		corrections that preserve conservation remain 
		underutilized in fluid domains despite their critical 
		role in FSI analysis, especially for the 
		multi-resolution scheme where fluid domains generally 
		have a low resolution.
		In this study, we incorporate the reverse kernel 
		gradient correction (RKGC) formulation, a 
		conservative high-order consistency approximation, 
		into the fluid discretization for solving FSI problems.
		RKGC has been proven to achieve exact second-order 
		convergence with relaxed particles and improve 
		numerical accuracy while particularly enhancing 
		energy conservation in free-surface flow simulations.
		By integrating this correction into the Riemann SPH 
		method to solve different typical FSI problems with
		a multi-resolution scheme, numerical results 
		consistently show improvements in accuracy and 
		convergence compared to uncorrected fluid discretization. 
		Despite these advances, further refinement of correction 
		techniques for solid domains and fluid-structure 
		interfaces remains significant for enhancing the overall 
		accuracy of SPH-based FSI modeling and simulation.
	\end{abstract}	
\end{frontmatter}
%
%
\section{Introduction}\label{introduction}
Fluid-structure interaction (FSI), a classical multiphysics 
problem characterized by the bidirectional coupling between 
fluids and deformable structures, is ubiquitous in nature, 
as seen in fish swimming, jellyfish pulsation, and insect 
flight, and critical to engineering applications such as 
wave-ship impacts, aircraft wing flutter, and blood-arterial 
wall dynamics.
Understanding FSI is fundamental to designing and optimizing 
these systems.
However, the inherent nonlinearity, time-dependent behavior, 
and moving fluid-structure interfaces pose significant 
numerical challenges, particularly when large deformations 
and free surfaces are involved~\cite{dowell2001modeling, 
	bungartz2006fluid, liu2019smoothed}.

Grid-based methods like the finite element method (FEM)
\cite{tezduyar1992new} have dominated numerical computations 
and are widely applied for solving FSI problems.
Techniques such as the immersed boundary method (IBM)
\cite{peskin2002immersed}, volume of fluid (VOF)
\cite{hirt1981volume}, and level set methods
\cite{osher1988fronts} are often employed to track moving 
interfaces and free surfaces.
Despite their widespread use, these methods face challenges 
in capturing sharp interfaces, ensuring mass conservation, 
and managing computational costs during frequent mesh 
updates for large structural deformations.

Lagrangian mesh-free methods, including smoothed particle 
hydrodynamics (SPH)~\cite{lucy1977numerical, 
gingold1977smoothed}, the discrete element method (DEM) 
\cite{mishra1992discrete}, and the moving particle 
semi-implicit method (MPS)~\cite{koshizuka1996moving}, 
offer advantages for FSI problems due to their advantages 
in handling free surfaces, large deformations, and 
material interfaces.
Hybrid approaches like SPH-FEM~\cite{yang2012free, 
chen2022multi, hu2014fluid, li2015non} and MPS-FEM 
\cite{zhang2019mps, chen2019numerical} combine 
Lagrangian particle methods for fluid modeling with 
FEM for structural analysis. 
While these methods improve numerical results, they 
still require careful treatment of fluid-structure 
interface boundary conditions.
Fully Lagrangian particle methods, such as SPH-SPH 
\cite{oger2009simulations, liu2013numerical, han2018sph, 
sun2019study, zhang2021multi, meng2022hydroelastic}, 
ISPH-SPH~\cite{khayyer2018enhanced, khayyer2021multi}, 
and MPS-MPS~\cite{hwang2016numerical, khayyer2019multi}, 
provide consistent modeling for both fluid and solid 
domains but numerical accuracy and stability remain 
key challenges for these Lagrangian particle methods.

To improve SPH consistency, methods like kernel gradient 
correction (KGC)~\cite{randles1996smoothed}, reproducing 
kernel particle methods (RKPM)~\cite{liu1995reproducing, 
liu1995reproducing2}, and corrective smoothed particle 
methods (CSPM)~\cite{chen1999corrective}, the finite 
particle method \cite{liu2006restoring}, modified SPH 
(MSPH)~\cite{batra2004analysis, zhang2004modified, 
sibilla2015algorithm, nasar2021high}, moving least squares 
(MLS)~\cite{atluri1999analysis}, and other high-order 
methods~\cite{flyer2011radial, king2020high, trask2017high} 
have been proposed.
These techniques enhance the accuracy of gradient and 
Laplacian operator approximations.
KGC, in particular, has been widely used for modeling 
solid structures~\cite{vignjevic2006sph, han2018sph, 
zhang2021multi} within the total Lagrangian framework, 
where it is used for discreting both momentum equations 
and deformation gradient tensors, demonstrating 
robustness in large-deformation analyses.
For fluid simulations, various KGC-liked corrections 
have been introduced to improve consistency while 
preserving conservation, such as averaged correction 
matrices~\cite{vila2005sph, zago2021overcoming, 
liang2023study} and pairwise particle corrections
\cite{oger2007improved, huang2022development}, etc. 
However, these approaches fail to achieve true 
high-order consistency~\cite{zhang2025towards}.

Despite these attempts and advancements, key challenges 
remain.
One significant difficulty is the trade-off between 
high-order consistency and momentum conservation 
\cite{zhang2025towards}.
While momentum conservation is important for numerical 
accuracy, most existing KGC implementations only rely 
on non-conservative formulations, limiting their 
applicability to systems governed by strict conservation 
laws.
Another challenge arises in the correction implementation 
of free-surface flows involving violent phenomena such 
as dam breaks~\cite{xue2022novel} and sloshing flows 
\cite{khayyer2021multi}, where rapid fluid fragmentation 
and extreme interface deformations can lead to 
ill-conditioned correction matrices, numerical errors, 
and simulation instabilities. 
Accurate fluid predictions are essential in FSI modeling 
since fluid dynamics errors propagate directly to the 
fluid-structure coupling, particularly in multi-resolution 
schemes where the fluid generally has a low resolution. 
Improving fluid accuracy enhances overall fidelity and 
ensures reliable predictions of complex interactions, 
such as force transmission and structural response.

In this study, we integrate the reverse kernel gradient 
correction (RKGC) method into the fluid domain 
discretization when solving FSI problems.
RKGC has been shown to satisfy first-order consistency 
in conservative formulations when proper particle 
relaxation is achieved and demonstrates improved 
accuracy in fluid simulations and fluid-rigid 
interactions~\cite{zhang2025towards}.
Building on this framework, both fluid and elastic solid 
domains are solved using the Riemann SPH method, where 
Riemann solutions govern particle-pair interactions. 
To ensure high-order consistency, non-dissipative terms 
in the fluid discretization are replaced with the RKGC 
formulation. 
The correction matrix for the fluid flow is updated at 
each advection time step, with additional weighted 
treatments for free-surface regions to maintain numerical 
stability.  
A multi-resolution scheme with larger fluid particle 
spacing balances accuracy and computational efficiency.
Numerical results consistently demonstrate that the 
proposed RKGC-corrected Riemann SPH method significantly 
improves accuracy in FSI simulations of both internal 
and free surface flows.

The remainder of this paper is organized as follows.
Section~\ref{numericalmethod} presents the SPH 
discretization for fluid, structure, and fluid-structure 
interactions, along with details of the RKGC Riemann SPH 
method.
Section~\ref{numericalexamples} validates the method 
through extensive numerical examples.
Section~\ref{conclusion} summarizes the key findings and 
suggests future research directions.    
%
%
\section{Numerical method}\label{numericalmethod}
\subsection{Fluid model}\label{fluidmodel}
The governing equations for an isothermal Newtonian fluid 
in the updated Lagrangian framework are the mass and 
momentum conservation equations, expressed as
\begin{equation}
	\begin{cases}
		\vspace{5pt}
		\displaystyle\dfrac{\text{d}\rho}{\text{d}t} 
		=-\rho\nabla\cdot\boldsymbol{\rm v}\\
		\displaystyle\dfrac{\text{d}\boldsymbol{\rm v}}
		{\text{d}t}=-\dfrac{1}{\rho}\nabla p+\boldsymbol{
			\rm a}_{\rm v}+\boldsymbol{\rm g}+\boldsymbol{
			\rm a}^{S:F},
	\end{cases}
	\label{wcsphgoverningequation}
\end{equation}
where $\rho$ is the fluid density, $t$ the time, $\boldsymbol
{\rm v}$ the velocity, and $p$ the pressure.
The term $\boldsymbol{\rm a}_{\rm v}=\nu\nabla^{2}\boldsymbol{
\rm v}$  represents the viscous acceleration with $\nu$ the 
kinematic viscosity, $\boldsymbol{\rm g}$ the gravitational 
acceleration, and 
$\boldsymbol{\rm a}^{S:F}=\boldsymbol{\rm f}^{S:F}/m$ 
denotes the interaction acceleration acting on the fluid 
due to the solid structure, where $\boldsymbol{\rm f}^{S:F}$ 
comprises both viscous and pressure forces and $m$ is the 
mass of the particle.
In addition, $\text{d}(\bullet)/\text{d}t=\partial(\bullet)/
\partial t+\boldsymbol{\rm v}\cdot\nabla(\bullet)$ refers 
to the material derivative.
To close the governing equations, an artificial equation 
of state (EoS) for weakly compressible flows is applied:
\begin{equation}
	p= {c_0}^2\left(\rho-\rho_{0}\right).
	\label{wcspheos}
\end{equation}
Here, $\rho_{0}$ is the initial density, and $c_{0}$ 
denotes the artificial sound speed.
Setting $c_{0}=10 U_{\rm max}$, where $U_{\rm max}$ 
represents the anticipated maximum fluid speed, satisfies 
the weakly compressible assumption where the density 
variation remains around 1\% \cite{morris1997modeling}.

The Riemann-SPH method \cite{zhang2017weakly}, which 
predicts inter-particle interactions via the Riemann 
solver, is employed to mitigate pressure oscillations 
in weakly compressible SPH (WCSPH).
Additionally, a multi-resolution SPH method is adopted 
to enhance computational efficiency \cite{zhang2021multi, 
	khayyer2021multi, chen2022multi, zhang2023efficient}.
Subsequently, the continuity and momentum equations are 
discretized as
\begin{equation}
	\begin{cases}
		\vspace{5pt}
		\displaystyle\dfrac{\text{d}\rho_{i}}{\text{d}t}
		=2\rho_{i}\sum_{j}\left(\boldsymbol{\rm v}_i-
		\boldsymbol{\rm v^*}\right)\cdot \nabla 
		W_{ij}^{h^{F}}V_{j}\\
		\displaystyle\dfrac{\text{d}\boldsymbol{\rm v}_{i}}
		{\text{d}t}= -\dfrac{2}{m_{i}}\sum_{j}P^{*}\nabla 
		W_{ij}^{h^{F}}V_{i}V_{j}+2\sum_{j}\dfrac{\nu}
		{\rho_{i}}\frac{\boldsymbol{\rm v}_{ij}}{r_{ij}}
		\dfrac{\partial W_{ij}^{h^{F}}}{\partial{r_{ij}}}
		V_{j}+\boldsymbol{\rm g}_{i}+
		\boldsymbol{\rm a}_{i}^{S:F}
		\left(h^{F}\right),
	\end{cases}
	\label{wcsphformulation}
\end{equation}
where $V_{i}$ is the particle volume, $\nabla W_{ij}^{h^F}=
\nabla W\left(\boldsymbol{\rm r}_{ij}, h^{F}\right)=\frac{
\boldsymbol{\rm e}_{ij}}{r_{ij}}\frac{\partial W_{ij}}{
\partial r_{ij}}$, where $\boldsymbol{\rm r}_{ij}=\boldsymbol{
\rm r}_{i}-\boldsymbol{\rm r}_{j}$ and $h^{F}$ is the 
smoothing length of the fluid, the unit vector 
$\boldsymbol{\rm e}_{ij}=\frac{\boldsymbol{\rm r}_{ij}}
{r_{ij}}$, denoting the derivative of the kernel function 
with respect to $\boldsymbol{\rm{r}}_i$, the position of 
the particle $i$.
The particle-pair velocity $\boldsymbol{\rm v}^{*}$ and 
pressure $P^{*}$ are solutions obtained from the Riemann 
problem constructed along the interacting line of each pair 
of particles \cite{zhang2017weakly, zhang2022smoothed}, 
where the left and right states of the Riemann problem 
are defined as
\begin{equation}
	\begin{cases}
		\left(\rho_{L}, U_{L}, P_{L}\right)=\left(\rho_{i}, 
		\boldsymbol{\rm v}_{i}\cdot\boldsymbol{\rm e}_{ij},
		p_{i}\right)\\
		\left(\rho_{R}, U_{R}, P_{R}\right)=\left(\rho_{j},
		\boldsymbol{\rm v}_{j}\cdot\boldsymbol{\rm e}_{ij},
		p_{j}\right).
	\end{cases}
\end{equation}
Using a linearized Riemann solver, the solutions for 
intermediate velocity and pressure are given by
\begin{equation}
	\begin{cases}
		\vspace{5pt}
		\boldsymbol{\rm v}^{*}=\overline{\boldsymbol{
				\rm v}}_{ij}+\left(U^{*}-\overline{U}_{ij}
		\right)\boldsymbol{\rm e}_{ij}\quad\\U^{*} = 
		\overline{U}_{ij} + \dfrac{1}{2}\dfrac{(p_i -
			p_j)}{\rho_0 c_0}\\
		P^{*}=\overline{p}_{ij}+
		\dfrac{1}{2}\beta\rho_{0}c_{0}U_{ij}.
	\end{cases}
	\label{linearsolution}
\end{equation}
Here, $\overline{(\bullet)}_{ij}=\left[(\bullet)_{i}
+(\bullet)_{j}\right]/2$ denotes the particle-pair average, 
$\overline{U}_{ij}=-\overline{\boldsymbol{\rm v}}_{ij}\cdot
\boldsymbol{\rm e}_{ij}$ and $U_{ij}=-\boldsymbol{\rm v}_{ij}
\cdot\boldsymbol{\rm e}_{ij}$, represents the particle-pair 
average and difference of the particle velocity along the 
interaction line, respectively.
The low-dissipation limiter, defined as $\beta=\min\left(
3\max\left(U_{ij}/c_{0},0\right),1\right)$, is introduced 
to reduce the inherent numerical dissipation and enhance
the discretization accuracy.
\subsection{Structure model}\label{structuremodel}
The solid structure is considered an elastic and weakly 
compressible material.
The governing equations in the total Lagrangian framework 
read as
\begin{equation}
	\begin{cases}
		\vspace{5pt}
		\displaystyle\rho^{S}=\left(\rho^{S}\right)^{0}J^{-1}\\
		\left(\displaystyle\dfrac{\text{d}\boldsymbol{\rm 
				v}}{\text{d}t}\right)^{S}=\dfrac{1}{\rho^{0}}
		\nabla^{0}\cdot \mathbb{P}^{T}+\boldsymbol{\rm g}
		+\boldsymbol{\rm a}^{F:S},
		\label{solidgoverningequation}
	\end{cases}
\end{equation}
where $\rho^{S}$ is the solid density, $\left(\rho^{S}
\right)^{0}$ is the initial reference density, and
$J={\rm det}\left(\mathbb{F}\right)$ denotes the Jacobian 
determinant of the deformation gradient tensor $\mathbb{F}$.
The operator $\nabla^{0}\left(\bullet\right)\equiv\partial\left
(\bullet \right)/\partial\boldsymbol{\rm r}^{0}$ represents 
the gradient with respect to the initial reference 
configuration. 
The superscript $\left(\bullet\right)^{0}$ indicates 
quantities in the initial reference state.
$\mathbb{P}^{T}$ is the nominal stress tensor, which is the 
transpose of the first Piola–Kirchhoff stress tensor $\mathbb{P}$.
Additionally, $\boldsymbol{\rm a}^{F:S}$ represents the 
acceleration induced by fluid-structure interaction.

For a Kirchhoff material, the first Piola–Kirchhoff stress 
tensor is given by
\begin{equation}
	\mathbb{P}=\mathbb{F}\mathbb{S},
\end{equation}
where the deformation gradient tensor $\mathbb{F}$ reads
\begin{equation}
	\mathbb{F}=\nabla^{0}\boldsymbol{\rm u}+\mathbb{I},
\end{equation}
with $\boldsymbol{\rm u}=\boldsymbol{\rm r}-\boldsymbol{
\rm r}^{0}$ representing the displacement and $\mathbb{I}$ 
the identity matrix.
The second Piola–Kirchhoff stress tensor $\mathbb{S}$ is 
related to the Green–Lagrange strain tensor $\mathbb{E}$ 
by the constitutive relation
\begin{equation}
	\displaystyle \mathbb{E}=\dfrac{1}{2}\left(
	\mathbb{F}^{T}\mathbb{F}-\mathbb{I}\right).
\end{equation}
For a linear elastic and isotropic material, the constitutive 
equation reads
\begin{align}
	\displaystyle\mathbb{S}&=K{\rm tr}\left(\mathbb{E}\right)
	\mathbb{I}+2G\left(\mathbb{E}-\dfrac{1}{3}{\rm tr}\left(
	\mathbb{E}\right)\mathbb{I}\right)\notag\\
	&=\lambda{\rm tr}\left(\mathbb{E}\right)\mathbb{I}+2\mu
	\mathbb{E},
\end{align}
where $\lambda$ and $\mu$ are the Lamé parameters 
\cite{sokolnikoff1946mathematical}.
The bulk modulus is given by $K=\lambda+\left(2\mu/3\right)$, 
and the shear modulus is $G=\mu$.
These parameters are related to the Young’s modulus $E$ and 
the Poisson ratio $\nu$ through
\begin{equation}
	E=2G\left(1+2\nu\right)=3K\left(1-2\nu\right).
\end{equation}
For weakly compressible materials, the speed of sound in the 
solid structure is defined as $c^{S}=\sqrt{K/\rho^{S}}$.

The discretization of the governing equations in Eq.
\eqref{solidgoverningequation} at particle $a$ is expressed 
as
\begin{equation}
	\begin{cases}
		\vspace{5pt}
		\displaystyle \left(\rho^{S}\right)_{a}=\left(\rho^{S}
		\right)^{0}_a{\rm det}\left(\mathbb{F}_{a}\right)^{-1}\\
		\displaystyle\left(\dfrac{\rm d \boldsymbol{{\rm v}}}
		{\rm d t}\right)^{S}_{a}=\dfrac{1}{m_{a}}\sum_{b}
		\left(\mathbb{P}_{a}\mathbf{B}^{0}_{a}+\mathbb{P}_{b}
		\mathbf{B}^{0}_{b}\right)\nabla^{0}W^{h^S}_{ab}V_{a}
		V_{b}+\boldsymbol{\rm g}_{a}+\boldsymbol{\rm a}_
		{a}^{F:S}\left(h^{S}\right),
	\end{cases}
\end{equation}
with the deformation gradient tensor discretized as
\begin{equation}
	\displaystyle\mathbb{F}_{a}=\left(\sum_{b}\left(
	\boldsymbol{\rm u}_b-\boldsymbol{\rm u}_a\right)\otimes
	\nabla^{0}W_{ab}^{h^{S}}V_{b}\right)\mathbf{B}^{0}_{a}+
	\mathbb{I},
\end{equation}
where $h^{S}$ is the smoothing length for the solid structure.
The correction matrix $\mathbf{B}$ \cite{randles1996smoothed, 
vignjevic2006sph}, which mitigates kernel inconsistency, is 
computed in the initial reference configuration as
\begin{equation}
	\displaystyle\mathbf{B}^{0}_{a}=\left(\sum_{b}\left(
	\boldsymbol{\rm r}^{0}_{b}-\boldsymbol{\rm r}^{0}_{a}
	\right)\otimes\nabla^{0}W^{h^{S}}_{ab}V_{b}^{0}\right)^{-1}.
	\label{solidcorrectionmatrix}
\end{equation}
\subsection{Fluid–structure coupling}
In fluid-structure interaction, structure particles are 
treated as a moving boundary for fluid particles, providing 
necessary boundary conditions for solving the momentum 
and continuity equations.
To ensure proper interaction, the smoothing length $h^{F}$ is 
chosen such that $h^F \geq h^{S}$, allowing a solid particle 
$a$ to be recognized and tagged as a neighbor of a fluid 
particle $i$.
The interaction acceleration exerted by the structure on the 
fluid is given by
\begin{equation}
	\displaystyle\boldsymbol{a}_{i}^{S:F}=-\dfrac{2}{m_{i}}
	\sum_{a}P^{*}\nabla W^{h^{F}}_{ia}V_{i}V_{a}+2\sum_{a}
	\dfrac{\nu}{\rho_{i}}\dfrac{\boldsymbol{\rm v}_{i}-
		\boldsymbol{\rm v}_{a}^{d}}{r_{ia}}\dfrac{\partial 
		W_{ia}^{h^{F}}}{\partial r_{ia}}V_{a},
\end{equation}
where the first and second terms on the right-hand side 
represent the accelerations induced by fluid pressure and 
viscous forces, respectively.
Here, $i$ refers to a fluid particle, and $a$ denotes a 
structure particle. The pressure term $P^{*}$ is obtained 
from a one-sided Riemann-based scheme \cite{zhang2023efficient}, 
where the left and right states are defined as
\begin{equation}
	\begin{cases}
		\left(\rho_{L}, U_{L}, P_{L}\right)=\left(\rho_{i}, 
		-\boldsymbol{\rm v}_{i}\cdot\boldsymbol{\rm n}_{a},
		p_{i}\right)\\
		\left(\rho_{R}, U_{R}, P_{R}\right)=\left(\rho_{a},-
		\boldsymbol{\rm v}_{a}^{d}\cdot\boldsymbol{\rm 
			n}_{a}, p_{a}^{d}\right),
	\end{cases}
\end{equation}
where $\boldsymbol{\rm n}_{a}$ is the local normal vector 
pointing from the solid to the fluid.
The imaginary pressure $p^{d}{a}$ and velocity 
$\boldsymbol{\rm v}_{a}^{d}$ are derived based on the no-slip 
boundary condition at the fluid–structure interface as
\begin{equation}
	\begin{cases}
		\vspace{5pt}
		\displaystyle p^{d}_{a}=p_{i}+\rho_{i}{\rm max}
		\left(0,\left(\boldsymbol{\rm g}-\frac{\widetilde{
				\rm d \boldsymbol{\rm v}_{a}}}{\rm dt}\right)
		\right)\cdot\boldsymbol{\rm r}_{ia}\\
		\boldsymbol{\rm v}_{a}^{d}=2\boldsymbol{\rm v}_{i}-
		\boldsymbol{\rm \widetilde{v}}_{a}.
	\end{cases}
\end{equation}
Here, $\frac{\widetilde{\rm d \boldsymbol{\rm v}_{a}}}{\rm dt}$ 
and $\boldsymbol{\rm \widetilde{v}}_{a}$ represent the 
time-averaged acceleration and velocity of solid particles 
over one fluid acoustic time step~\cite{zhang2020dual} to 
resolve the mismatch between force and acceleration when 
using the multi-resolution scheme~\cite{zhang2021multi, 
	zhang2023efficient}. 
Consequently, the interaction acceleration 
$\boldsymbol{a}_{i}^{F:S}$ exerted by the fluid on the solid 
structure can be determined in a straightforward manner.
\subsection{RKGC Riemann SPH and transport velocity formulation}
For a smooth field $\psi\left(\boldsymbol{\rm r}\right)$, 
the SPH approximation of its gradient using the reverse kernel 
gradient correction (RKGC)~\cite{zhang2025towards} at a 
particle position $\boldsymbol{\rm r}_{i}$ can be expressed as
\begin{equation}
	\displaystyle\nabla\psi_{i}=-\sum_{j}\left(
	\psi_{i}\mathbf{B}_{j}+\psi_{j}\mathbf{B}_{i}
	\right)\nabla W_{ij} V_{j},
	\label{rkgccorrection}
\end{equation}
where the KGC matrix is applied in reverse order with respect 
to particles \(i\) and \(j\), and \(\mathbf{B}_{i}\) is the 
correction matrix \cite{randles1996smoothed}, defined as 
\begin{equation}
	\mathbf{B}_{i}=\left(-\sum_{j}\mathbf{r}_{ij}\otimes\nabla  
	W_{ij}V_{j}\right)^{-1}.
	\label{fluidcorrectionmatrix}
\end{equation}
The calculation in Eq.~\eqref{fluidcorrectionmatrix} follows 
the same formulation as in Eq.~\eqref{solidgoverningequation}, 
except that the current version for fluids is based on the 
updated configuration and calculated in each advection time 
step~\cite{zhang2020dual}.
The gradient approximation in Eq.~\eqref{rkgccorrection} has 
been shown to exactly satisfy first-order consistency when 
particle relaxations are based on $\mathbf{B}_{i}$ and 
$\mathbf{B}_{j}$, a form of “geometric stress” that depends 
on particle positions.
Accordingly, the original particle-pair average term in the 
Riemann solution, given in Eq.~\eqref{linearsolution} as \(\overline{(\bullet)}_{ij}=\left[(\bullet)_{i}+(\bullet)_{j}
\right]/2\), is modified in the momentum equation as
\begin{equation}
	\overline{p}_{ij}\Rightarrow\overline{\left(p\mathbf{B}
		\right)}_{ij}=\frac{1}{2}\left(p_{i}\mathbf{B}_{j}+p_{j}
	\mathbf{B}_{i}\right),
	\label{pressurecorrection}
\end{equation}
to improve pressure prediction accuracy in solving 
fluid-structure interaction (FSI) problems.
A previous study~\cite{zhang2025towards} demonstrated that 
applying the kernel gradient correction (KGC) in the continuity 
equation has negligible impact on simulation results for the 
Lagrangian SPH method.  
Following this observation, we adopt the same practice in the 
present study without correction on the continuity equation.

However, computing the correction matrix may lead to numerical 
instability, particularly in simulations of violent free-surface 
flows~\cite{zago2021overcoming}. 
For example, during wave breaking and splashing events, the 
correction matrix can become ill-conditioned, introducing 
significant errors and instability.
To address this issue, the weighted kernel gradient correction 
(WKGC)~\cite{ren2023efficient} is employed, where the KGC 
matrix near the free surface region is weighted with the 
identity matrix (WKGC$^1$) to improve numerical robustness.
In this approach, the weighted KGC matrix is defined as
\begin{equation}
	\widetilde{\mathbf{B}}_{i}=\omega_{1}\mathbf{B}_{i}
	+\omega_{2}\mathbf{I},
\end{equation}
where the weighting coefficients \(\omega_{1}\) and \(\omega_{2}\) 
are given by  
\begin{equation}
	\begin{cases}
		\vspace{5pt}
		\displaystyle\omega_{1}=\frac{\left|\mathbf{A}_{i}
			\right|}{\left|\mathbf{A}_{i}\right|+\kappa_{i}}\\
		\displaystyle\omega_{2}=\frac{\kappa_{i}}
		{\left|\mathbf{A}_{i}\right|+\kappa_{i}}.
	\end{cases}
\end{equation}
Here, \(\left|\mathbf{A}_{i}\right|=\left|\mathbf{B}_{i}^{-1}
\right|\) serves as a smoothness indicator for the correction 
matrix, while \(\kappa_{i}=\max\left(\alpha-\left|\mathbf{A}_{i}
\right|, 0\right)\) determines the weighting applied to the 
identity matrix.
The parameter \(\alpha\) acts as a threshold to regulate the 
weighting process.  
Further details on this approach can be found in Ref. 
\cite{ren2023efficient}.

Since particle relaxation remains computationally expensive in 
Lagrangian SPH methods \cite{zhu2021cad, zhang2025towards}.
To address this, a KGC-based transport velocity formulation is 
employed, which involves a single correction step per time step. 
The transport velocity $\widetilde{\boldsymbol{\rm v}}_{i}$   
governs particle position updates via
\begin{equation}
	\displaystyle\dfrac{\rm d\boldsymbol{\rm r}_i}{\rm {dt}}=
	\widetilde{\boldsymbol{\rm v}}_{i}.
\end{equation}
Numerically, this formulation applies a one-step correction to 
particle positions during each advection time 
step~\cite{zhang2020dual},
\begin{equation}
	\Delta\mathbf{r}_i = \eta\left(\Delta{x}\right)^2\sum_j
	(\mathbf{B}_{i}+\mathbf{B}_{j})\nabla W_{ij}V_j,
	\label{Brelaxation}
\end{equation}
where the parameter $\eta$ is generally chosen as 0.2 to 
ensures stability by aligning with the CFL number criteria 
in SPH~\cite{monaghan1992smoothed}.
Note that the transport velocity formulation is imposed on 
the internal flow but is not applied to free-surface flows 
due to the complexity of handling the free surface and 
general practices in SPH simulations~\cite{oger2007improved, 
ren2023efficient, lyu2023derivation, zhang2025towards}.       
%
%
%
\section{Numerical examples}\label{numericalexamples}
In this section, we present several typical numerical cases 
to validate the enhanced accuracy of the proposed RKGC 
Riemann SPH method for modeling and simulating FSI problems.
For all simulations, the C2 Wendland kernel is adopted with 
smoothing length $h^{F}=1.3dp^{F}$ and $h^{S}=1.15dp^{S}$, 
where $dp^{F}$ and $dp^{S}$ denote the initial particle 
spacing for the fluid and solid structure, respectively. 
In addition, according to the multi-resolution framework 
established in our prior work~\cite{zhang2021multi}, a 
resolution ratio of $dp^{F}=2dp^{S}$ is employed unless 
otherwise stated, which strategically reduces the 
computational cost by coarsening the fluid particle 
resolution while preserving simulation accuracy through 
targeted refinement at fluid-structure interfaces.
\subsection{Hydrostatic water column on an elastic plate}
\begin{figure}[htb]
	\centering
	\includegraphics[width=0.6\textwidth]
	{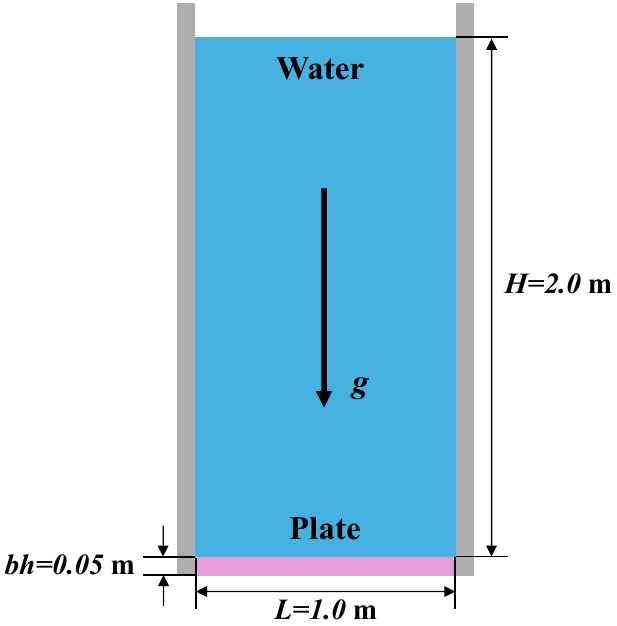}
	\caption{Hydrostatic water column on an 
		elastic plate: Schematic illustration.}
	\label{hydrostaticschematicsketch}
\end{figure}

The first example presents a typical benchmark test involving 
a hydrostatic water column resting on an elastic aluminum plate, 
originally proposed by Fourey et al.~\cite{fourey2017efficient}.
This case has been widely adopted for FSI validation studies 
\cite{zhang2021multi, khayyer2021multi, khayyer2023enhanced, 
	xue2022novel, bao2024entirely}.
As illustrated in Fig.~\ref{hydrostaticschematicsketch}, the 
water tank has dimensions of 1.0~m (width) × 2.0~m (height).
The tank's bottom consists of an aluminum plate with 
a thickness of $bh=0.05$~m, a density of 
$\rho^S=2700~\mathrm{kg/m^3}$, a Young’s modulus of 
$E^S=67.5~\mathrm{GPa}$, and a Poisson’s ratio of 
$\nu^S=0.34$.
At $t=0.0$~s, the plate is instantaneously subjected 
to the hydrostatic load.
After initial oscillations, the FSI system reaches an 
equilibrium state.
The theoretical equilibrium displacement at the plate's 
mid-span, derived from the analytical model in Ref. 
\cite{fourey2017efficient}, is $-6.85\times10^{-5}$~m.
Following established methodologies~\cite{fourey2017efficient, 
khayyer2018enhanced, zhang2021multi}, a constant fluid 
time-step size of $\Delta t^{F}=2\times 10^{-5}$~s is 
employed, and the simulations terminate at $t=2.0$~s.
The artificial damping method from Ref.~\cite{zhu2022dynamic} 
is also adopted to mitigate the oscillations.

\begin{figure}[htb!]
	\centering
	\includegraphics[width=0.6\textwidth]
	{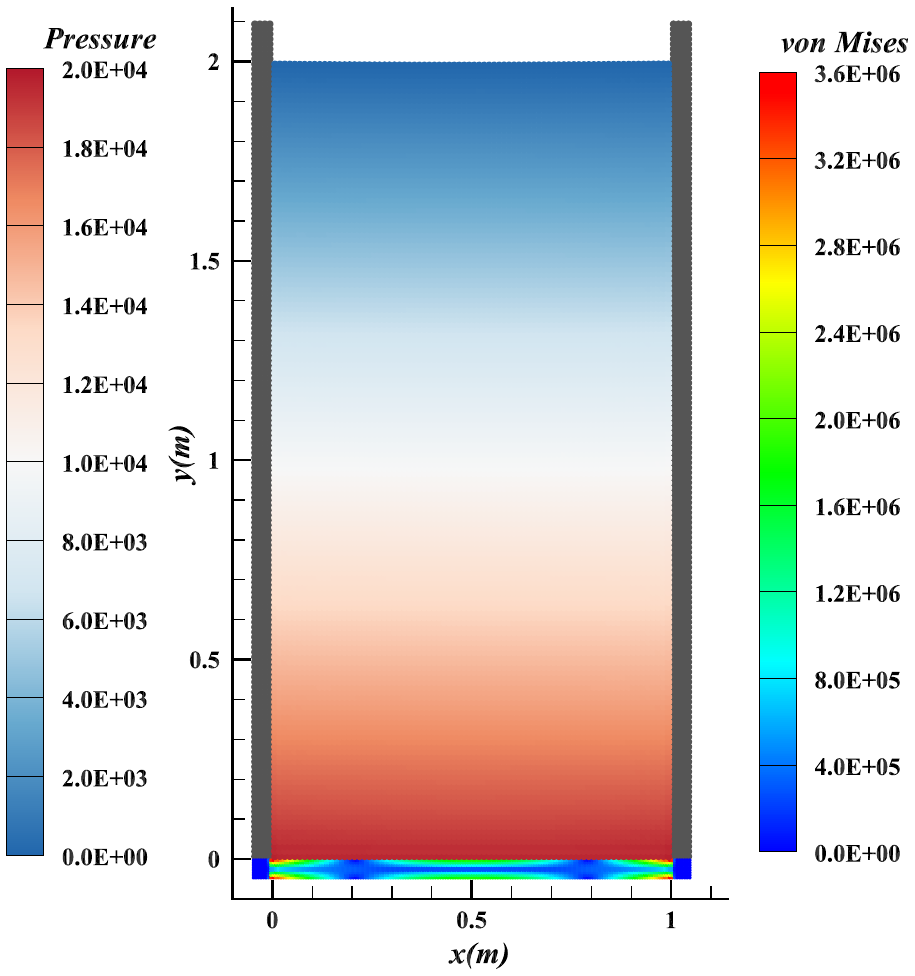}
	\caption{Hydrostatic water column on an elastic plate: 
		Distributions of fluid pressure and structural 
		von Mises stress fields at $t = 0.5~\mathrm{s}$.}
	\label{hydrostaticcontour}
\end{figure}

\begin{figure}[htb]
	\centering
	\includegraphics[width=\textwidth]
	{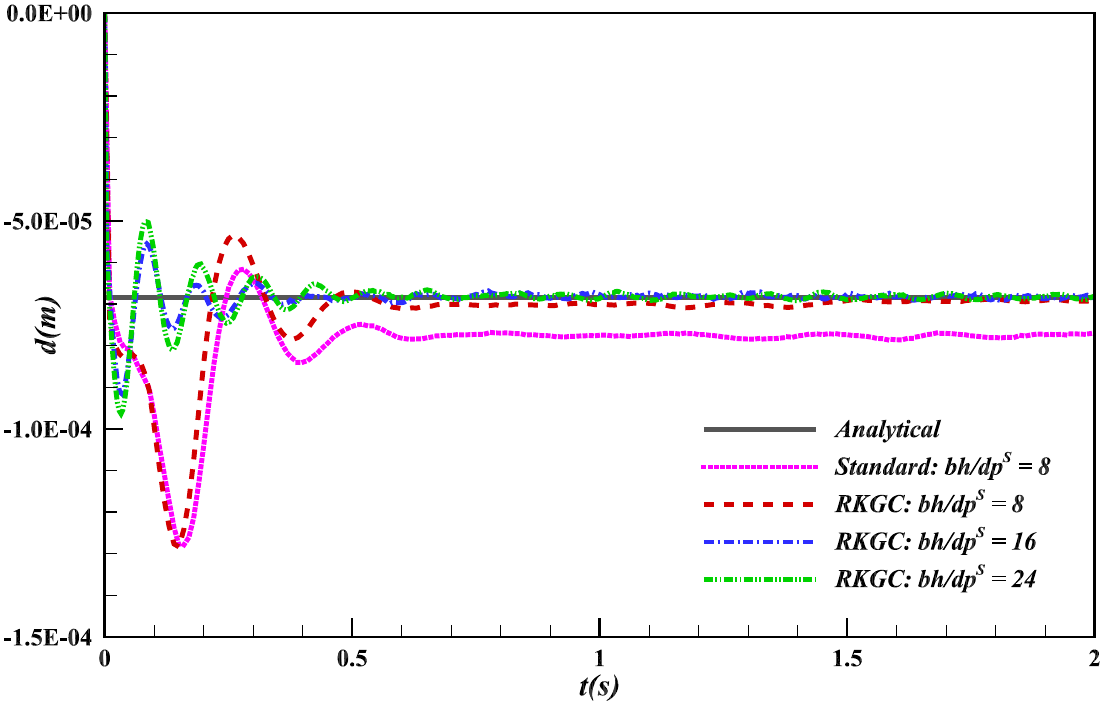}
	\caption{Hydrostatic water column on an elastic plate: 
		Time history of the vertical mid-span displacement 
		of the elastic plate.}
	\label{hydrostaticdisplacement}
\end{figure}

Fig.~\ref{hydrostaticcontour} shows fluid pressure and 
structural von Mises stress distributions at $t=0.5$~s 
for the RKGC method at the resolution $bh/dp^{S}=8$.
The present RKGC corrected method predicts smooth pressure 
and stress fields in the respective domains, aligning with 
the results reported in Refs.~\cite{khayyer2009enhanced, 
khayyer2021multi}, indicating good numerical stability. 
The time history of the vertical mid-span displacement of 
the elastic plate predicted by different methods, as well 
as for different resolutions of the present method, is 
presented in Fig.~\ref{hydrostaticdisplacement}.
The standard SPH method predicts a displacement of $-8.0\times 
10^{-5}$~m at the resolution $bh/dp^S = 8$, consistent with 
the findings of Zhang et al.~\cite{zhang2021multi} (as shown 
in their Fig.~3(b)) and Khayyer et al.~\cite{khayyer2018enhanced, 
khayyer2021multi} (as shown in their Fig.~8 at $bh/dp^S = 12$ 
and Fig.~6 at $bh/dp^S = 10$, respectively).
In contrast, the RKGC SPH method predicts a displacement 
at the same resolution that closely aligns with the 
analytical solution from Ref.~\cite{fourey2017efficient}.
This is because the pressure force from the water primarily 
drives the displacement of the aluminum plate, and the RKGC
method improves the accuracy of the pressure predictions. 
With the resolution increases, the results converge toward 
the analytical solution, demonstrating the good convergence 
of the proposed method.

\begin{figure}[htb]
	\centering
	\includegraphics[width=\textwidth]
	{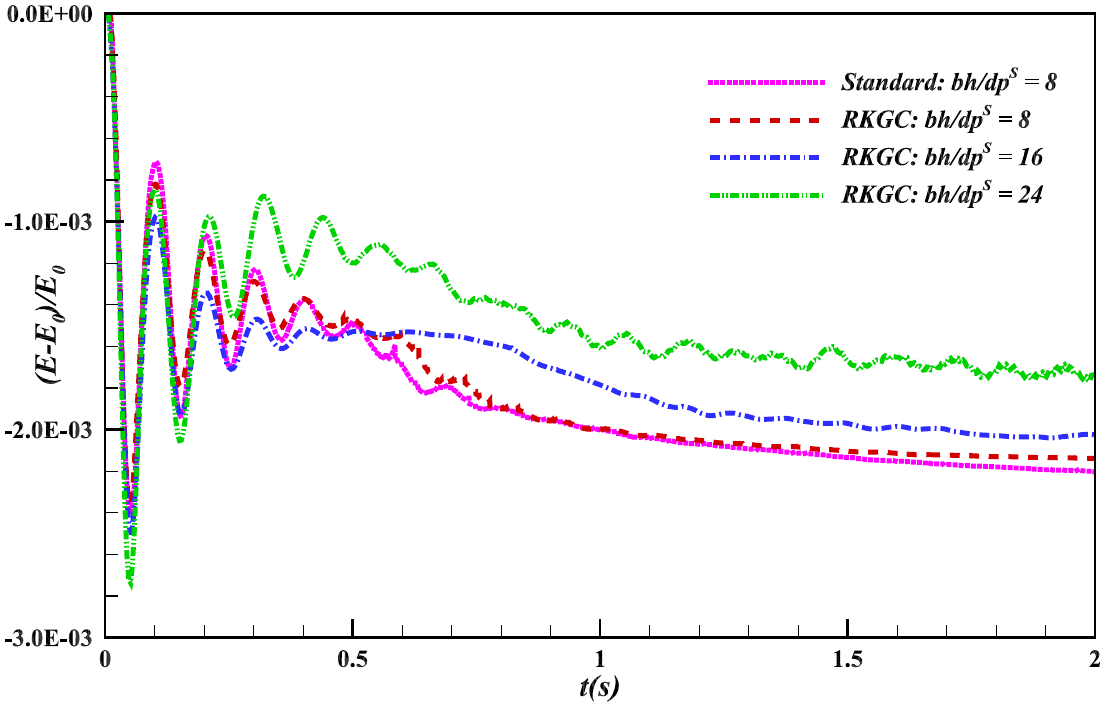}
	\caption{Hydrostatic water column on an elastic plate: 
		Time history of the normalized total energy for 
		different methods.}
	\label{hydrostaticenergy}
\end{figure}

To validate the energy conservation property of the present 
method, Fig.~\ref{hydrostaticenergy} illustrates the time 
history of the normalized total energy of the system computed 
using different methods, as well as for different resolutions 
of the present method.
The total energy drops at a rate between $-2.0 \times 10^{-3}$ 
and $-1.0 \times 10^{-3}$, which is consistent with the results 
reported in Refs.~\cite{zhang2021multi, khayyer2021multi}.
The RKGC SPH method exhibits a slightly lower rate of energy 
decay compared to the standard SPH method at the lowest 
resolution, demonstrating the enhanced energy conservation 
property of the proposed method. 
As the resolution increases, the energy decay rate decreases, 
indicating the improved energy conservation.
%
\subsection{Flow-induced vibration of a beam behind 
	a cylinder}
\begin{figure}[htbp]
	\centering
	\includegraphics[width=\textwidth]{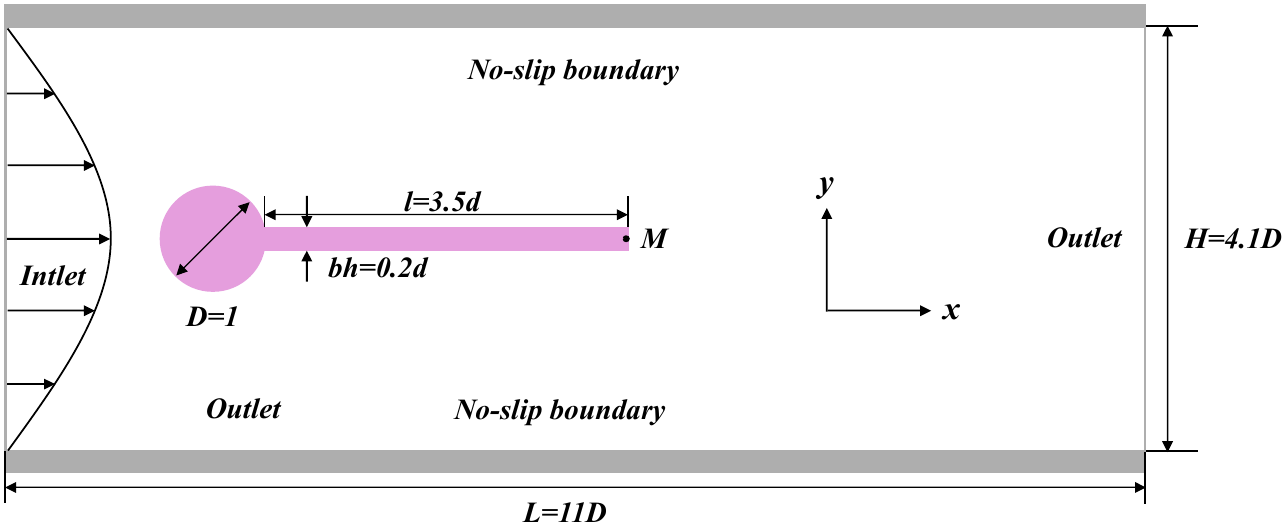}
	\caption{Flow-induced vibration of a beam behind a cylinder: 
	Schematic illustrate. Sensor $M$ monitors the trajectory 
	at the free end of the beam.}
	\label{fsi2configuration}	
\end{figure}

The second example involves two-dimensional flow-induced 
vibration (FIV) of a flexible beam attached to a rigid cylinder, 
corresponding to the widely studied FSI2 benchmark test 
proposed by Turek and Hron~\cite{turek2006proposal}.	
The computational setup, illustrated in Fig.~\ref{fsi2configuration}, 
features a rigid cylinder of diameter $D=1$ centered at $(2D, 2D)$ 
within a domain spanning $11D \times 4.1D$. 
A sensor, labeled $M$, tracks the trajectory at the free 
end of the beam.
No-slip boundary conditions are enforced on the top and 
bottom walls, while the left and right boundaries are designated 
as inflow and outflow, respectively. 
The inflow velocity follows a time-dependent parabolic profile 
given by
\begin{equation}
	U(y)=1.5\overline{U}(t,y) \frac{(H - y)y}{H^2},
\end{equation}
where
\begin{equation}
	\overline{U}(t,y) = 
	\begin{cases} 
		\vspace{5pt}
		0.5U_0\left(1.0-\cos\left(0.5\pi t\right)\right) 
		& \text{if } t < t_s, \\
		U_0 & \text{otherwise},
	\end{cases}
\end{equation}
with $U_0=1.0$ and $t_s=2.0$, and $H$ denoting the channel 
height.
Consistent with previous studies~\cite{turek2006proposal, 
han2018sph, zhang2021multi, sun2021accurate}, the physical 
parameters replicate the challenging low-stiffness FSI2 
benchmark test.
The density ratio of the structure to the fluid is $\rho^S/\rho^F=10$,
and the Reynolds number is $Re = \rho^FU_0D/\eta=100$.
The beam is modeled as an isotropic linear elastic material 
with a dimensionless Young’s modulus 
$E^{*}=E^{S}/\rho^F/U^2_0=1.4\times10^3$ and Poisson’s ratio 
$\nu^{S}=0.4$.

\begin{figure}[htbp]
	\vspace{-3.2cm}
	\centering
	\includegraphics[width=0.93\textwidth]{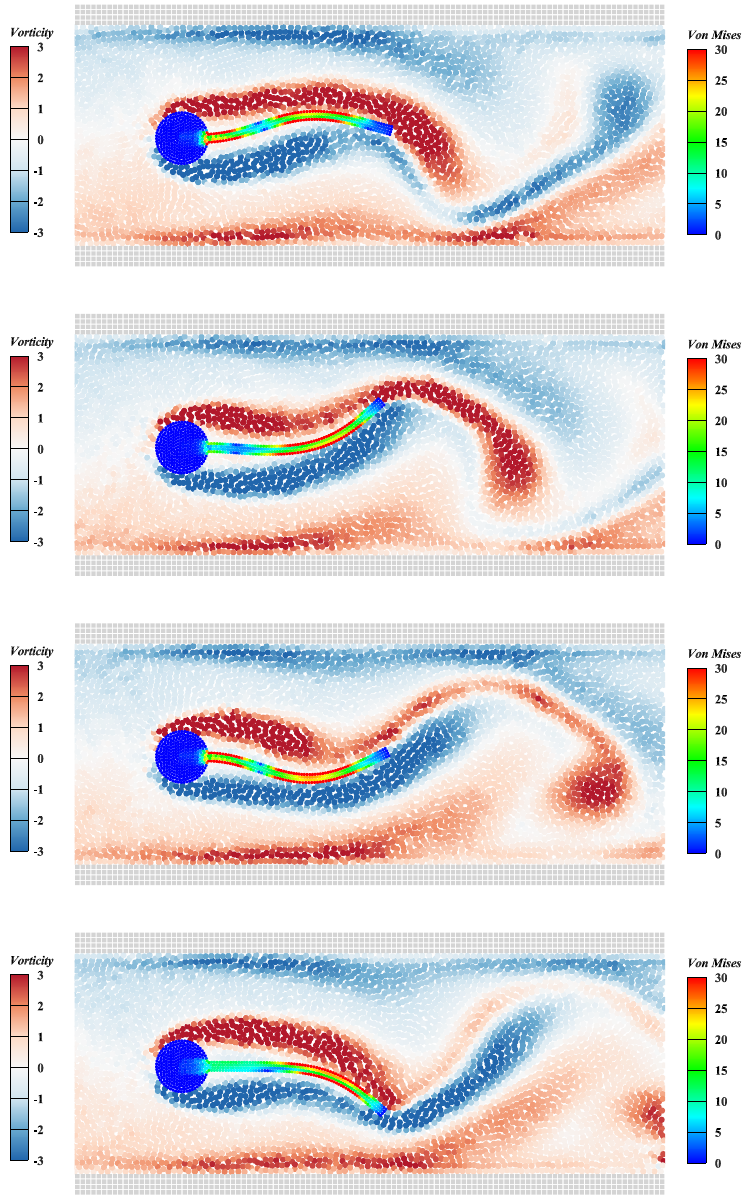}
	\caption{Flow-induced vibration of a beam behind a cylinder: 
		Distributions of the von Mises stress in the elastic
		plate and the vorticity in the flow field at four 
		time instants.}
	\label{fsi2contour}	
\end{figure}

Fig.~\ref{fsi2contour} shows the von Mises stress distribution 
in the elastic plate and the vorticity in the flow field at 
four time instants after self-sustained oscillations are 
established.
These results, obtained using the present RKGC corrected 
method at the resolution $bh/dp^s=4$, demonstrate clear 
periodic motion consistent with other numerical results from 
Refs. \cite{turek2006proposal, bhardwaj2012benchmarking, 
	han2018sph, zhang2021multi, zheng2024parameter}. 

\begin{figure}[htbp]
	\vspace{-3cm}
	\centering
	\includegraphics[width=0.93\textwidth]{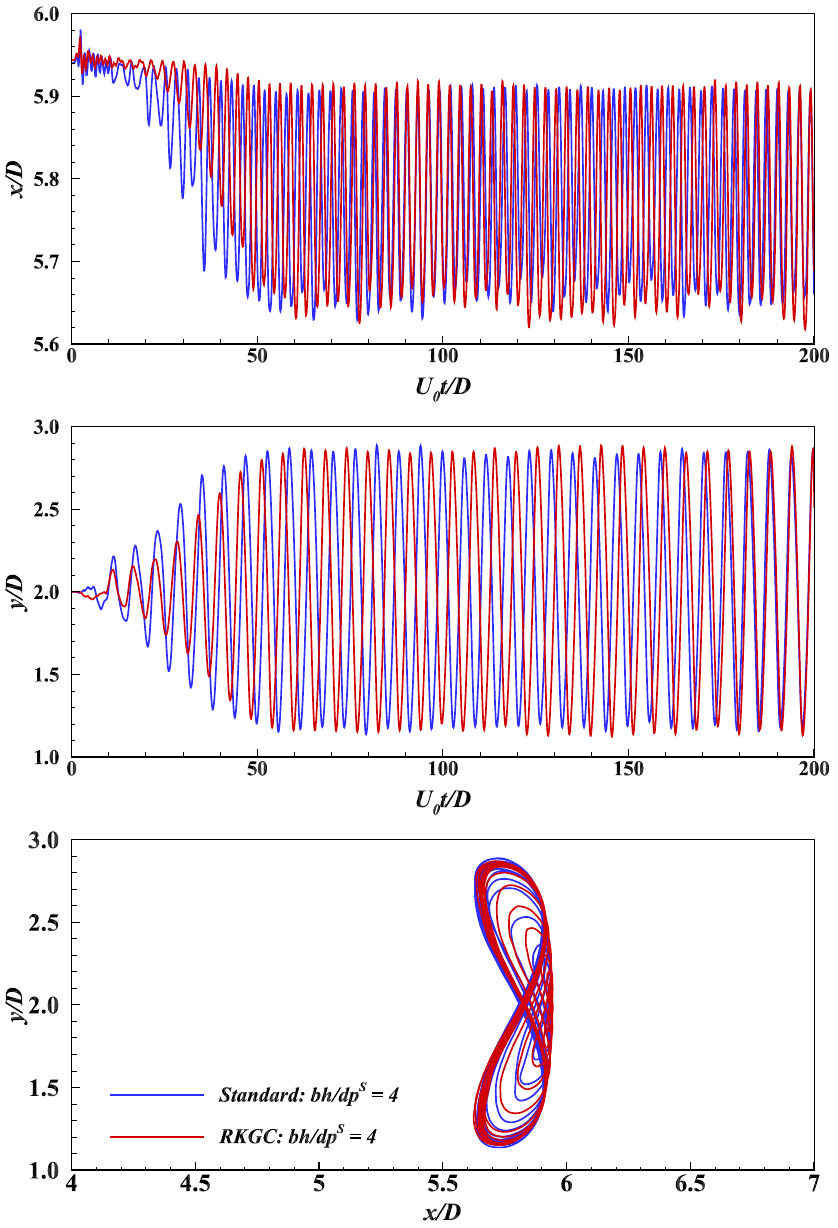}
	\caption{Flow-induced vibration of a beam behind a cylinder: 
		Amplitude of the displacement in x-direction (top panel), 
		amplitude of the displacement in y-direction (middle panel) 
		and the trajectory (bottom panel) of point $M$ obtained 
		by different methods.}
	\label{fsi2displacement}	
\end{figure}

Fig.~\ref{fsi2displacement} presents the displacement amplitudes 
in the x-direction and y-direction, along with the trajectory 
of point $M$, obtained by different methods at the resolution 
$bh/dp^s=4$.  
As time exceeds a dimensionless value of 50, the beam exhibits 
periodic self-sustained oscillations. 
The trajectory of point $M$ follows a typical Lissajous curve, 
with a 2:1 frequency ratio between horizontal and vertical 
motions~\cite{turek2006proposal}, indicating strong agreement 
with results reported in Refs.~\cite{turek2006proposal, 
	bhardwaj2012benchmarking, zhang2021multi}. 
It is evident that at this resolution, the amplitudes obtained 
by the two methods are similar, while the frequency for the 
present RKGC method is clearly higher.

\begin{figure}[htb!]
	\centering
	\includegraphics[width=1\textwidth]{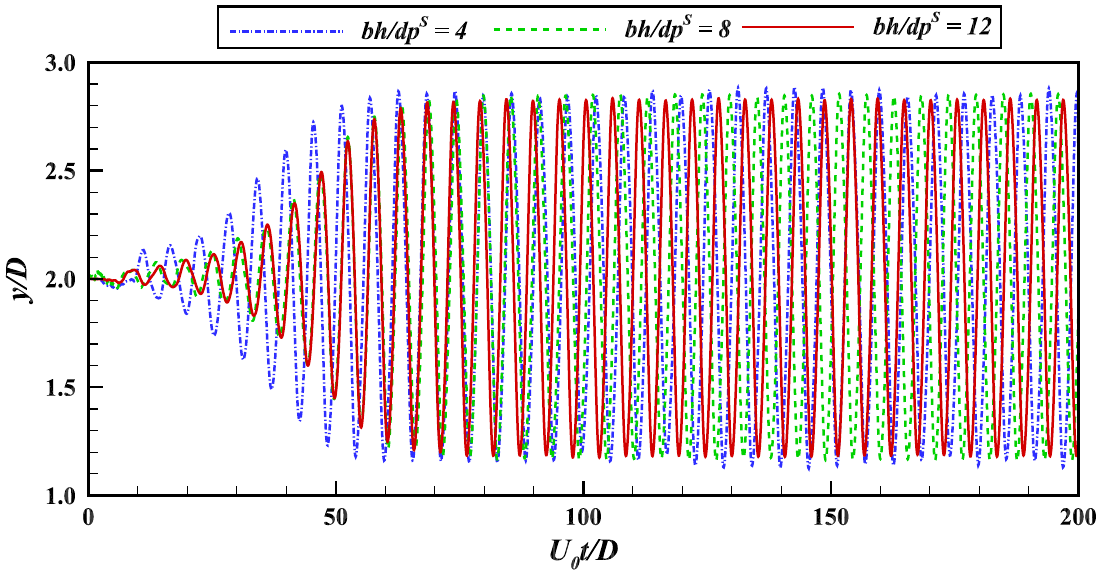}
	\caption{Flow-induced vibration of a beam behind 
		a cylinder: Convergence study of the RKGC method 
		on the amplitude of the displacement in y-direction.}
	\label{fsi2displacementrkgc}	
\end{figure}

To enable a quantitative comparison, 
Table~\ref{fsi2displacementcomparison} presents reference 
results, results obtained using different methods, and 
convergence studies of the present RKGC method.
\begin{table}[htb!]
	\small
	\renewcommand\arraystretch{1.25}
	\centering
	\captionsetup{font={small}}
	\caption{Flow-induced vibration of a beam behind a cylinder: 
		Comparison results with the reference and the convergence 
		study of the present RKGC method.}
	\begin{tabularx}{13.5cm}{@{\extracolsep{\fill}}ccccc}
		\hline
		\quad Results source &Amplitude in y-axis &Frequency \quad\\
		\midrule
		\quad Turek and Hron \cite{turek2006proposal}
		& $0.830$ & $0.190$ \quad\\ 
		\quad Bhardwaj and Mittal \cite{bhardwaj2012benchmarking}
		& $0.920$ & $0.190$ \quad\\
		\quad Tian et al. \cite{tian2014fluid}
		& $0.784$ & $0.190$ \quad\\
		\quad Han and Hu \cite{han2018sph}
		& $0.886$ & $0.168$ \quad\\
		\quad Zhang et al. \cite{zhang2021multi} 
		& $0.860$ & $0.188$ \quad\\
		\quad Zheng et al. \cite{zheng2024parameter} 
		& $1.090$ & $0.174$ \quad\\
		\quad RKGC($bh/dp^{S}=12$)
		& $0.831$ & $0.187$ \quad\\
		\quad RKGC($bh/dp^{S}=8$)
		& $0.833$ & $0.183$ \quad\\
		\quad RKGC($bh/dp^{S}=4$)
		& $0.861$ & $0.175$ \quad\\
		\quad Standard($bh/dp^{S}=4$)
		& $0.847$ & $0.171$ \quad\\
		\bottomrule
	\end{tabularx}
	\label{fsi2displacementcomparison}
\end{table}
At low resolution ($bh/dp^{S}=4$), the present method shows 
closer with reference amplitude and frequency values than the 
standard method, though both underestimate frequencies. 
Increasing resolution improves frequency predictions and reduces 
y-direction amplitude, converging toward literature values.
The convergence trend of the y-direction amplitude is shown in 
Fig.~\ref{fsi2displacementrkgc}, demonstrating good convergence.
%
\subsection{Dam-break flow through an elastic gate}
\label{dambreakthroughgate}
\begin{figure}[htb!]
	\centering
	\includegraphics[width=0.8\textwidth]{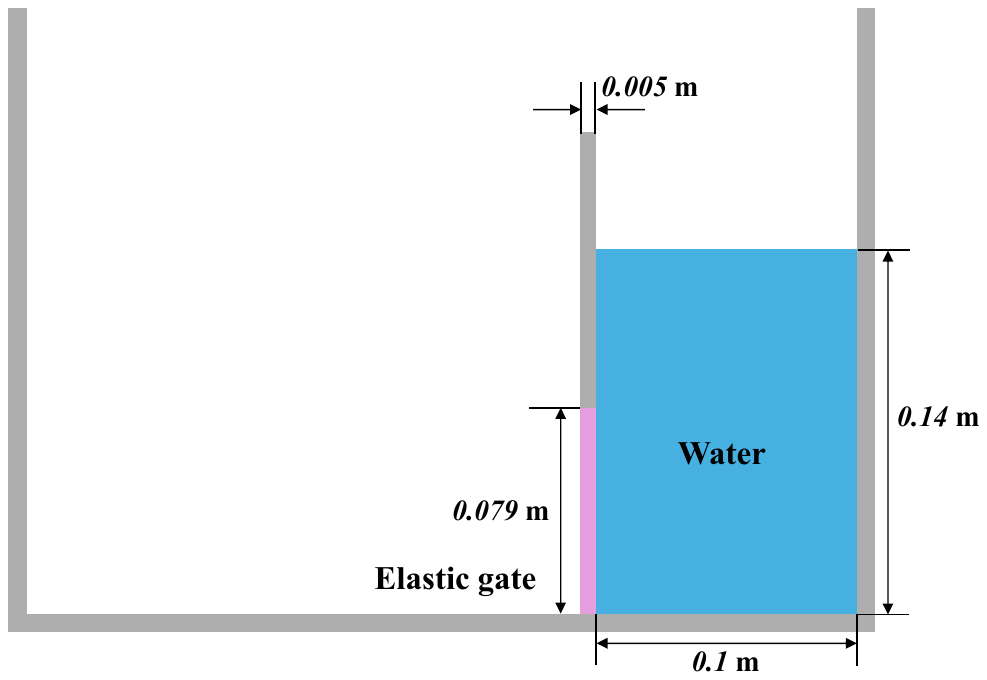}
	\caption{Dam-break flow through an elastic gate: Schematic 
		illustration.}
	\label{dambreakthroughgateconfiguration}	
\end{figure}

The third example involves the deformation of an elastic gate 
subjected to time-dependent water pressure with free-surface flow. 
As illustrated in Fig.~\ref{dambreakthroughgateconfiguration}, 
the gate is clamped at its upper end and free at the lower end, 
interacting with a body of water initially confined in an 
open-air tank behind it. 
This setup replicates the experimental study by Antoci et 
al.~\cite{antoci2007numerical}.
In accordance with Ref.~\cite{antoci2007numerical}, the fluid is 
treated as inviscid flow with a density of $\rho^F=1000~\mathrm{kg/m^3}$. 
The elastic gate, modeled as rubber, follows a linear isotropic 
material law with a density of $\rho^S=~1100\mathrm{kg/m^3}$, 
Young’s modulus $E^S=7.8~\mathrm{MPa}$, and Poisson’s ratio 
$\nu^S=0.4$. 
A clamped boundary condition is enforced at the upper end of the 
gate via a rigid base constraint.

\begin{figure}[htb!]
	\centering
	\includegraphics[width=\textwidth]
	{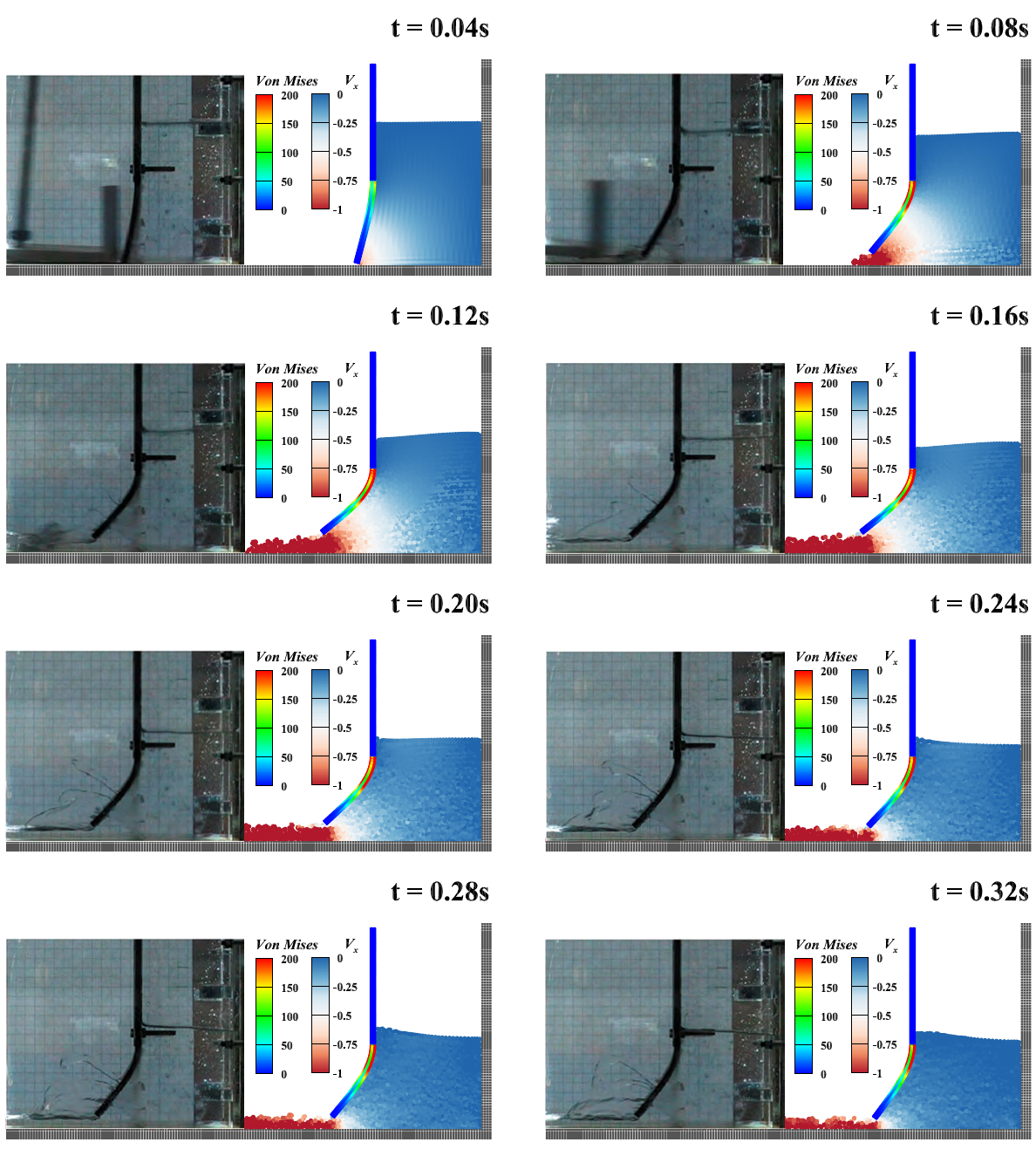}
	\caption{Dam-break flow through an elastic gate: Snapshots 
		from the RKGC method are compared with experimental 
		frames from Antoci et al. \cite{antoci2007numerical} at 
		eight equally spaced time instants.}
	\label{dambreakelasticgatecontour}
\end{figure}

Fig.~\ref{dambreakelasticgatecontour} compares snapshots from the 
present RKGC corrected method with experimental frames from Antoci et 
al.~\cite{antoci2007numerical} at eight equally spaced time instants. 
The simulation accurately captures the free-surface motion and 
closely matches the experimental results.
At $t=0.12$~s, the gate exhibits significant deformation, 
appearing more open than in the experiment.
Additionally, side-wall splashes observed experimentally from
$t=0.12$~s due to leakage between the tank wall and flexible 
gate~\cite{antoci2007numerical, rafiee2009sph} are not reproduced 
in this two-dimensional simulation.

\begin{figure}[htb!]
	\centering
	\includegraphics[width=1\textwidth]
	{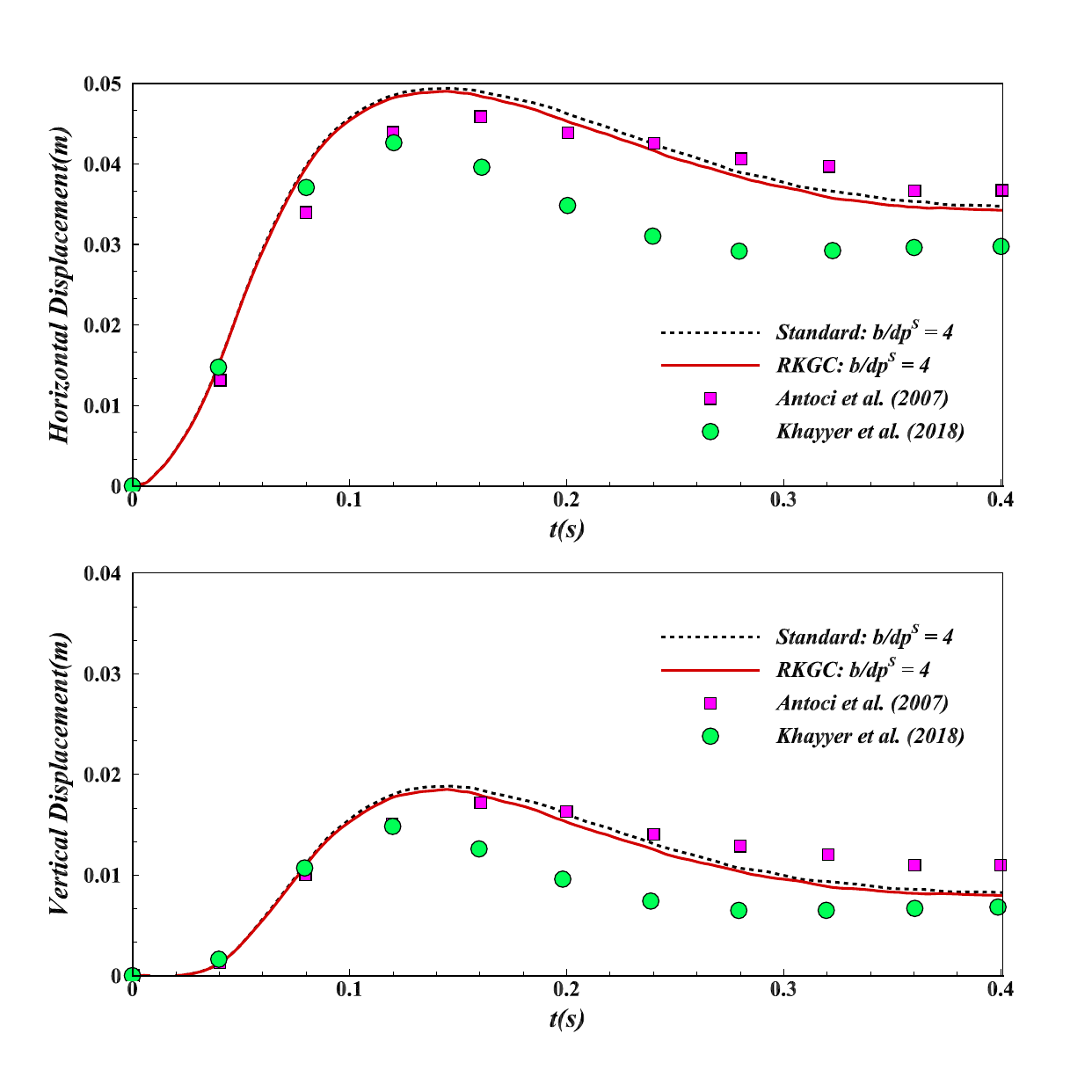}
	\caption{Dam-break flow through an elastic gate: 
		Horizontal (top panel) and vertical (bottom panel) 
		displacement of the free end of the gate obtained by 
		different methods.}
	\label{dambreakelasticgatedisplacement}
\end{figure}

Fig.~\ref{dambreakelasticgatedisplacement} compares horizontal and 
vertical displacements of the gate’s free end from the present 
simulations with experimental data~\cite{antoci2007numerical} and 
numerical results from Khayyer et al.~\cite{khayyer2018enhanced}, 
which used the same Poisson’s ratio ($\nu^{S}=0.4$).
Acceptable agreement is observed with both experimental and 
prior numerical results.
During the initial dam-break phase, the horizontal displacement 
increases rapidly due to high static pressure, with standard 
and RKGC SPH methods predicting similar gate behavior.
At $t=0.15$~s, the gate reaches maximum deformation, where 
the present RKGC method predicts a displacement closer to 
experimental measurements.
As water recedes and pressure diminishes, the gate gradually 
returns to equilibrium, governed by elastic forces and dynamic 
pressure.
It is also worth noting that, similar to previous simulations 
\cite{zhang2019smoothed, khayyer2018enhanced, zhang2021multi}, 
small differences are observed between the present results and 
the experimental data during the closing phase. 
These discrepancies can be attributed to the different material 
model used in simulations~\cite{yang2012free} and the fact that 
the selected Poisson’s ratio is lower than that of the actual 
rubber used in the experiment.
\begin{figure}[htb!]
	\centering
	\includegraphics[width=1\textwidth]
	{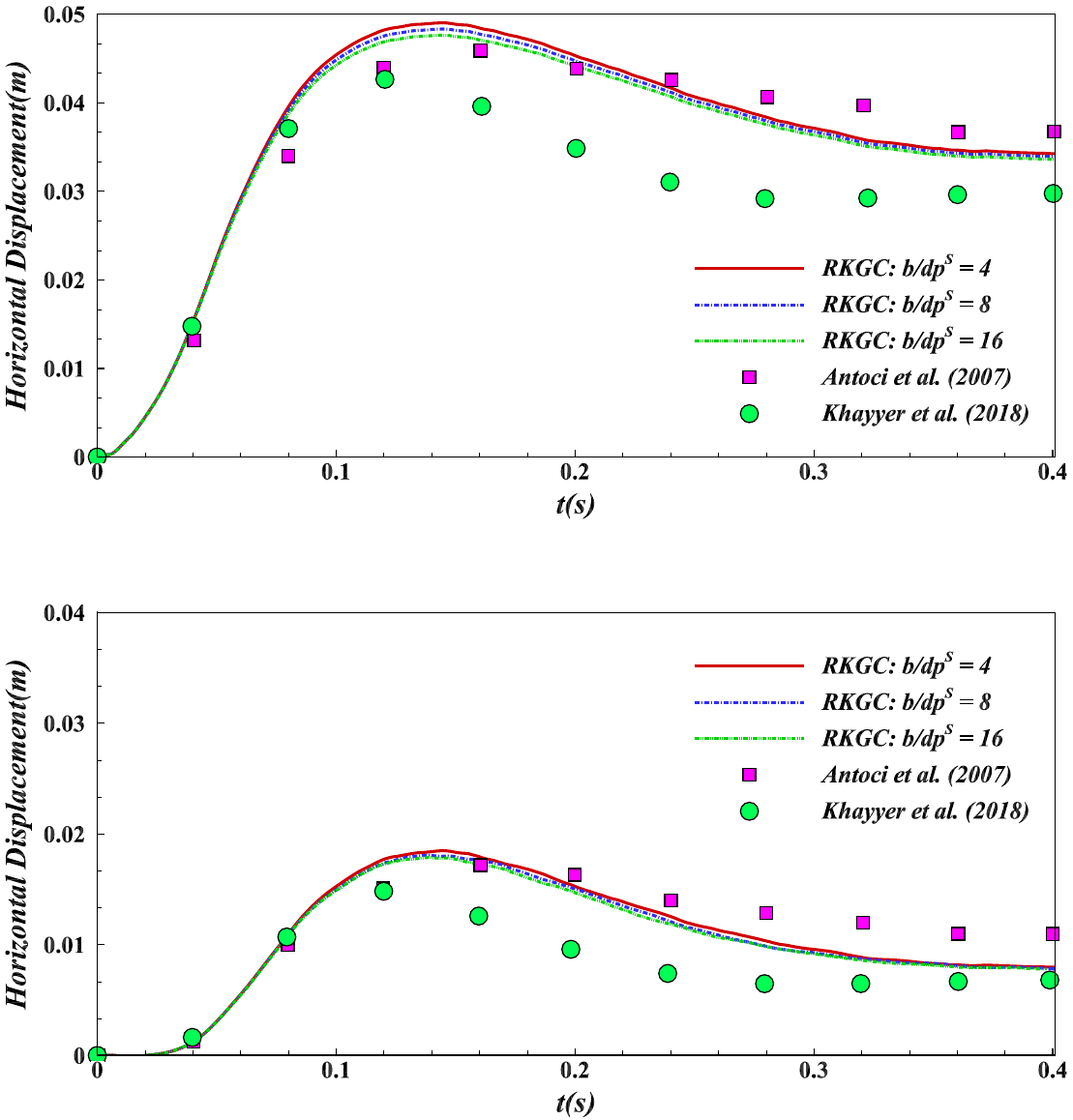}
	\caption{Dam-break flow through an elastic gate: 
		Convergence study of the RKGC method on horizontal 
		(top panel) and vertical (bottom panel) displacement 
		of the free end of the gate.}
	\label{dambreakelasticgateconvergence}
\end{figure}
Although the present method shows slightly larger deviations, 
convergence studies in Ref.~\cite{zhang2021multi} (as shown in 
their Fig. 11 with $\nu^S=0.47$), and that in 
Fig.~\ref{dambreakelasticgateconvergence} for the current case 
all indicate that the results obtained by the present RKGC 
method are closer to the converged solution during the gate 
closing, again confirming the improved numerical accuracy. 
%
\subsection{Dam-break flow impacts an elastic plate}
\begin{figure}[htb]
	\centering
	\includegraphics[width=0.8\textwidth]{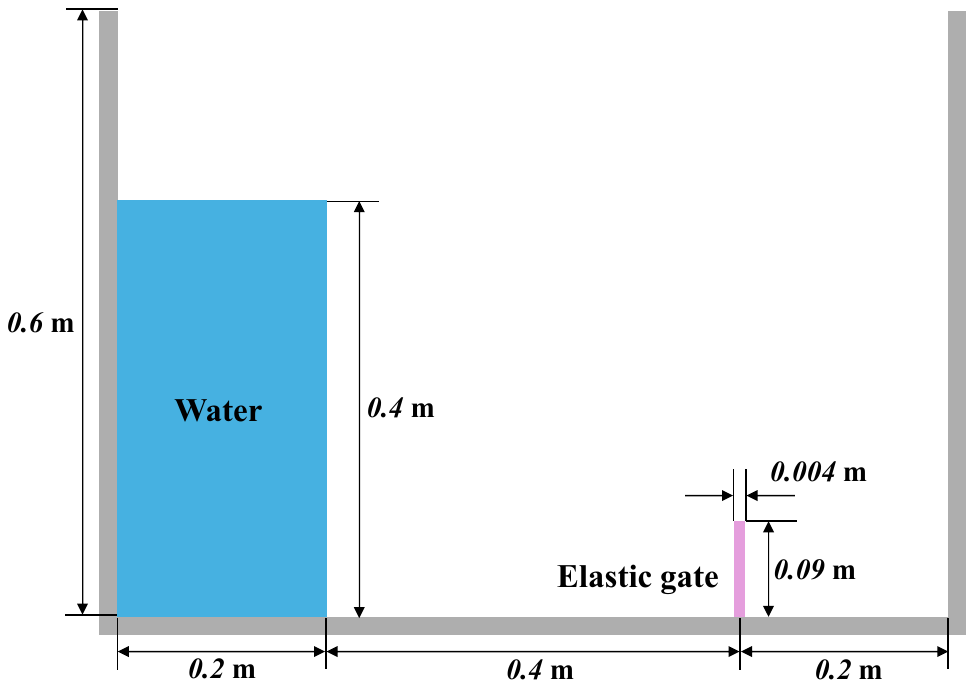}
	\caption{Dam-break flow impacts an elastic plate: 
		Schematic illustration.}
	\label{dambreakwithelasticgateconfiguration}	
\end{figure}

The fourth example examines a dam-break flow impacting an 
elastic plate, as experimentally investigated by Liao et 
al.~\cite{liao2015free}. 
This case involves violent fluid-structure interaction with 
structural deformations significantly larger than those in 
Section~\ref{dambreakthroughgate}, posing considerable 
computational challenges.
The computational setup, shown schematically in 
Fig.~\ref{dambreakwithelasticgateconfiguration}, aligns with 
the experimental configuration from Ref.~\cite{liao2015free}.
The water has a density of $\rho^F=998~\mathrm{kg/m^3}$ and 
an initial height of $H=0.4~\mathrm{m}$.
The gate material has a density of $\rho^S=1161.54~\mathrm{kg/m^3}$, 
Young's modulus of $E^S=3.5~\mathrm{MPa}$, Poisson's ratio of 
$\nu^S=0.49$, and a thickness of $b=0.004~\mathrm{m}$.  
The selected particle resolution for the simulation is $b/dp^S=4$.
Initially, a gate blocks the water; upon release, vertical motion 
induced by the system rapidly accelerates the gate, generating 
a dam-break flow that subsequently impacts the elastic plate.

\begin{figure}[htb!]
	\centering
	\makebox[\textwidth][c]{
		\subfigure[]{\includegraphics[width=0.45\textwidth]
			{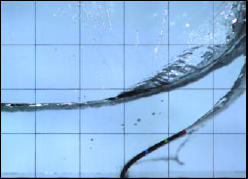}}}
	\makebox[\textwidth][c]{
		\subfigure[]{\includegraphics[width=.5\textwidth]
			{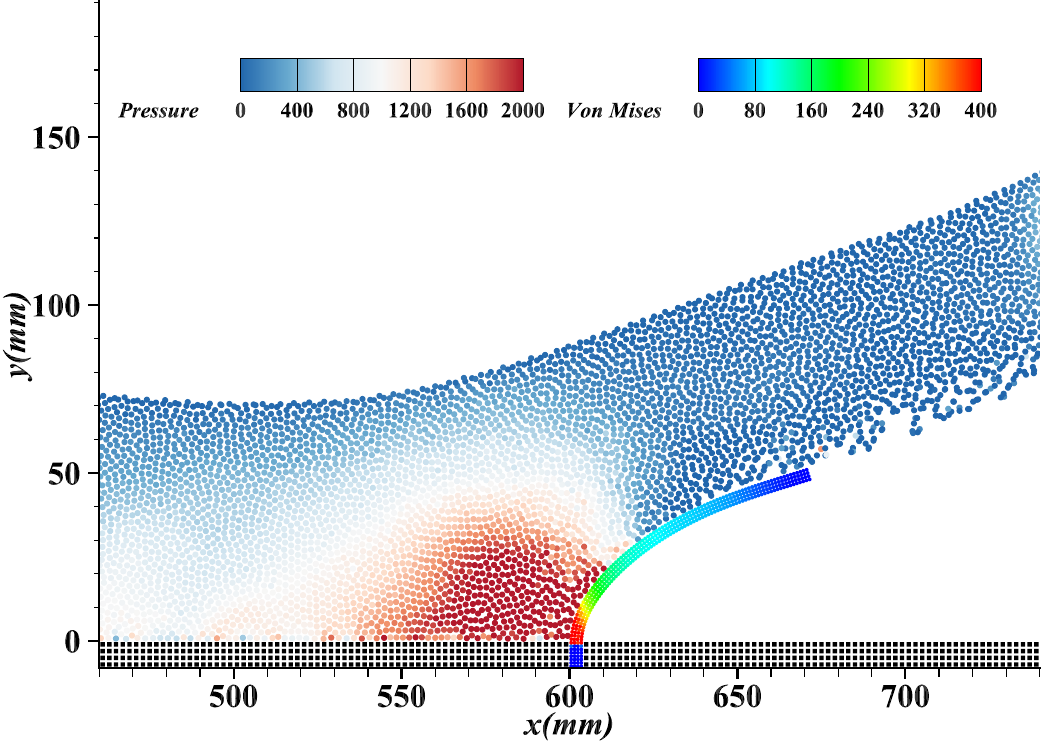}}
		\subfigure[]{\includegraphics[width=.5\textwidth]
			{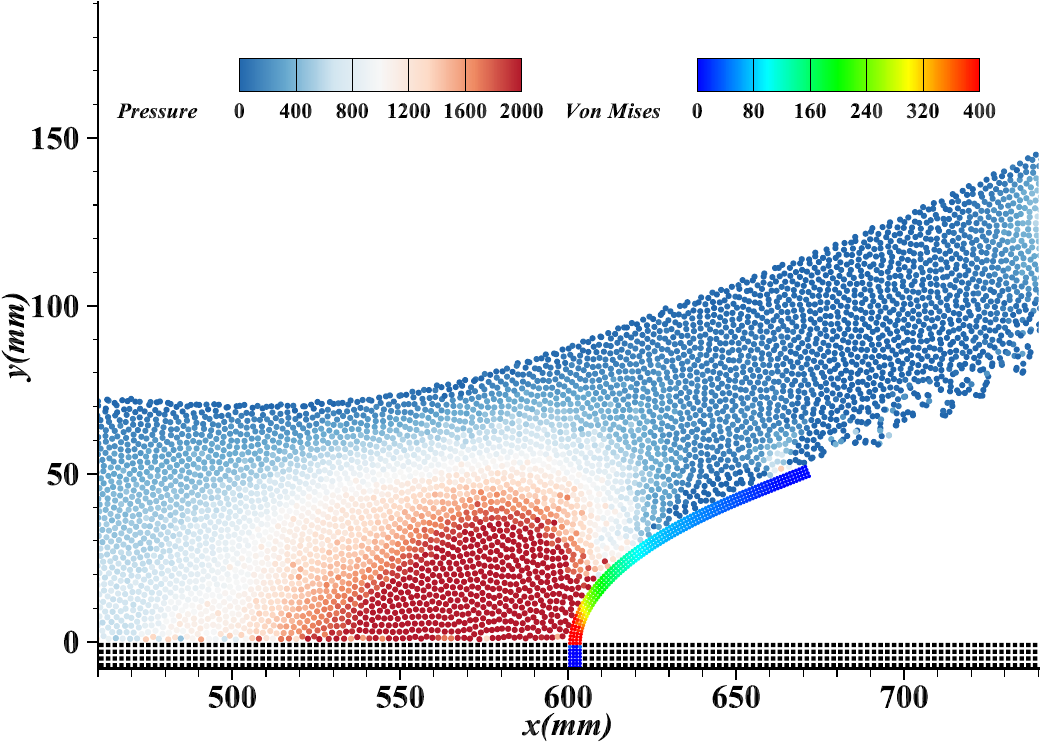}}}
	\caption{Dam-break flow impacts an elastic plate: 
		Comparison of the pressure distribution in the flow 
		field and deformation of the elastic plate from 
		(a) experimental snapshot captured by Liao et al. 
		\cite{liao2015free} and obtained by the (b) Standard 
		SPH method and by the (c) RKGC SPH method at $t=0.35$~s.} 
	\label{dambreakimpactplatecontour}
\end{figure}

Fig.~\ref{dambreakimpactplatecontour} presents the pressure 
distribution in the flow field and the von Mises stress of the 
elastic plate obtained using different methods, compared with 
an experimental snapshot from Liao et al.~\cite{liao2015free} 
at $t=0.35$~s.
Both the standard SPH and RKGC SPH methods produce smooth 
pressure and stress fields while maintaining stable free 
surfaces and structural deformations. 
However, the RKGC SPH method offers superior accuracy, 
particularly in pressure predictions at the base of the plate. 
This aligns with the findings of Khayyer et al.~\cite{khayyer2021multi} 
(as shown in their Fig.~18) and Xue et al.~\cite{xue2022novel} 
(as shown in their Fig.~19), where higher pressures were 
observed in this region. 
In contrast, the standard SPH method tends to underestimate 
pressure, highlighting the improved predictive capability 
of the RKGC approach.

\begin{figure}[htbp]
	\vspace{-2cm}
	\centering
	\includegraphics[width=1\textwidth]
	{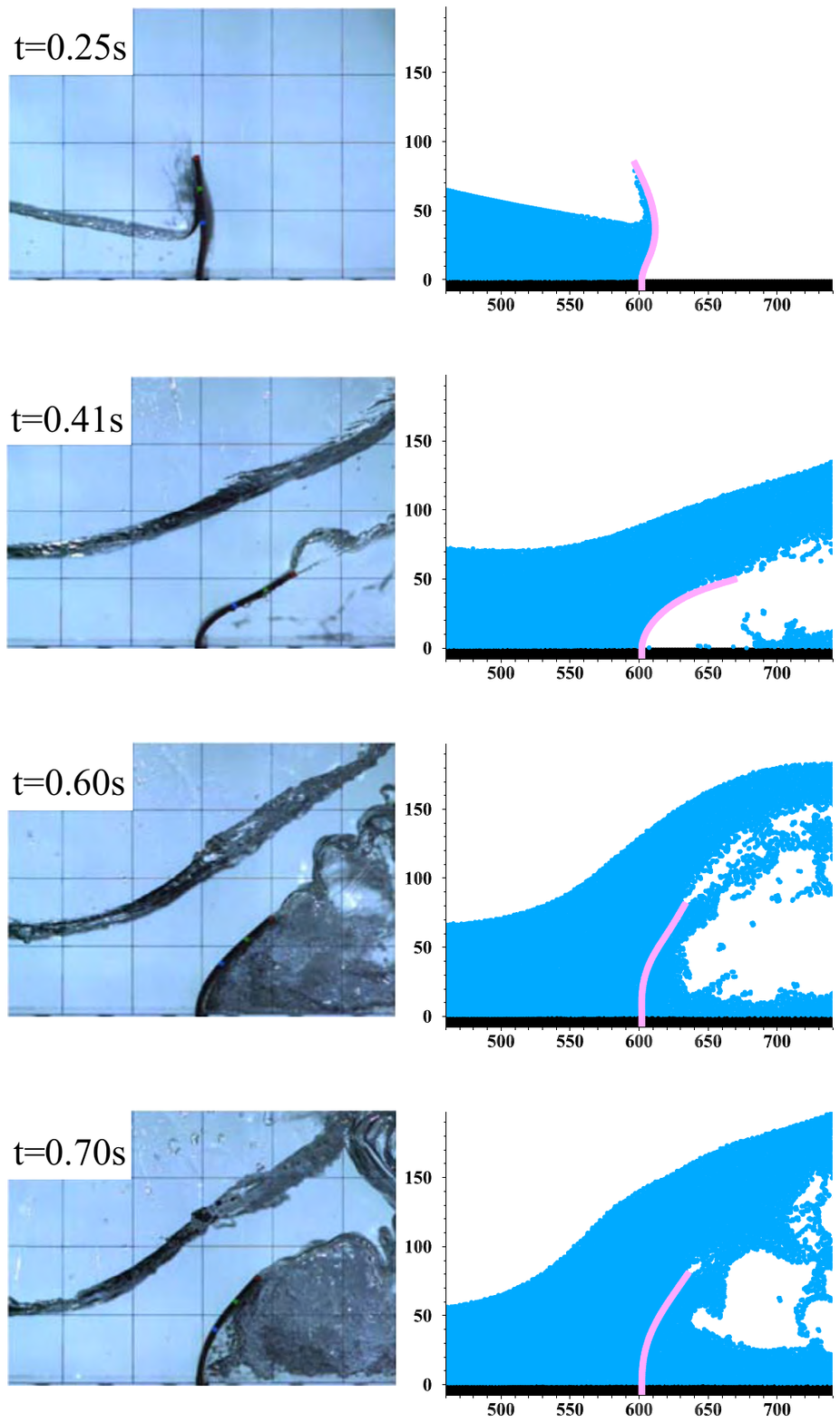}
	\caption{Dam-break flow impacts an elastic plate: 
		Comparison between simulation results obtained by the 
		RKGC method and experimental results~\cite{liao2015free} 
		at four typical time instants.}
	\label{dambreakimpactplateother}
\end{figure}

Fig.~\ref{dambreakimpactplateother} compares results from the 
RKGC method with experimental data at four key time instants.
Overall, the predicted free surface and elastic plate 
deformation align well with experimental observations. 
The experiment reveals that as the fluid impacts the opposite 
side of the elastic plate, air entrainment leads to significant 
water spray and cavity formation. 
This phenomenon is replicated in the simulation, resulting 
in violent particle splashing. 
However, some discrepancies arise in the later stages, where 
fluid-structure interactions become increasingly intense.

\begin{figure}[htb!]
	\centering
	\includegraphics[width=1\textwidth]
	{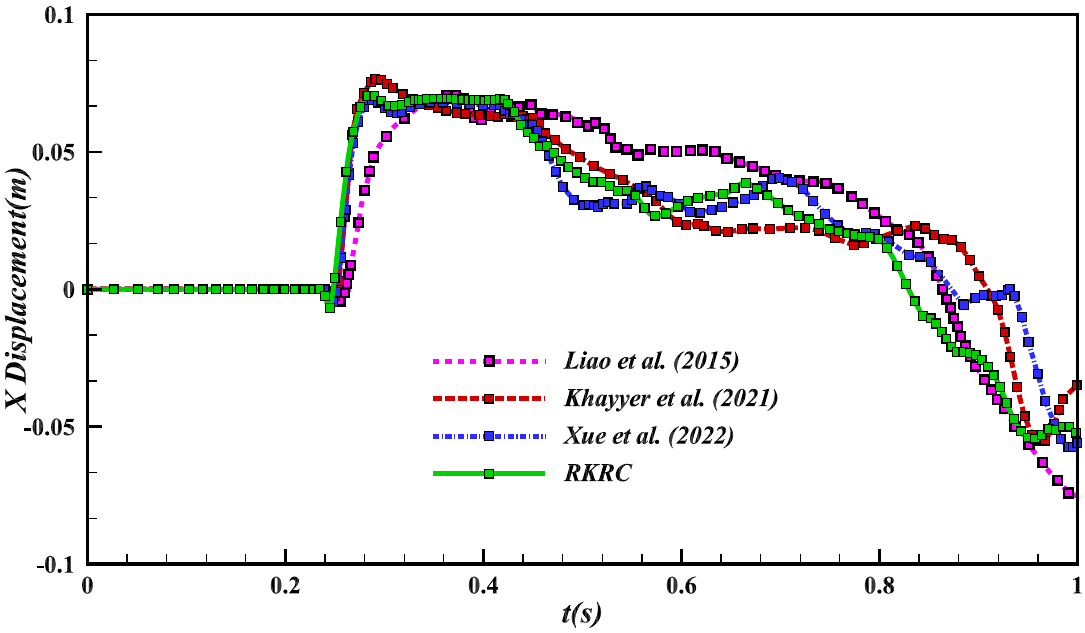}
	\caption{Dam-break flow impacts an elastic plate: 
		Time history of horizontal displacement of 
		free end of the elastic plate.}
	\label{dambreakelasticplateobservation}
\end{figure}

Fig.~\ref{dambreakelasticplateobservation} illustrates the 
time history of the free-end displacement of the elastic 
plate predicted by the present RKGC method and other reference 
predictions, compared with experimental measurements.
Similar to other numerical results reported in Refs.   
\cite{khayyer2021multi, xue2022novel}, the RKGC method 
closely follows experimental trends. 
However, as noted in Fig.~\ref{dambreakimpactplateother}, 
discrepancies persist in later stages due to the absence 
of air effects in the current water-structure interaction 
model. 
Upon impact, the water jet entraps air, forming cavities 
that significantly influence the flow field and structural 
response. 
This limitation is well-documented Refs.~\cite{khayyer2021multi, 
sun2021accurate}, suggesting that incorporating multi-phase 
air-water-structure simulations would enhance the accuracy 
of deformation predictions in future studies.
%
\subsection{Sloshing in a rolling tank with a elastic baffle}
\begin{figure}[htbp]
	\centering
	\includegraphics[width=0.8\textwidth]{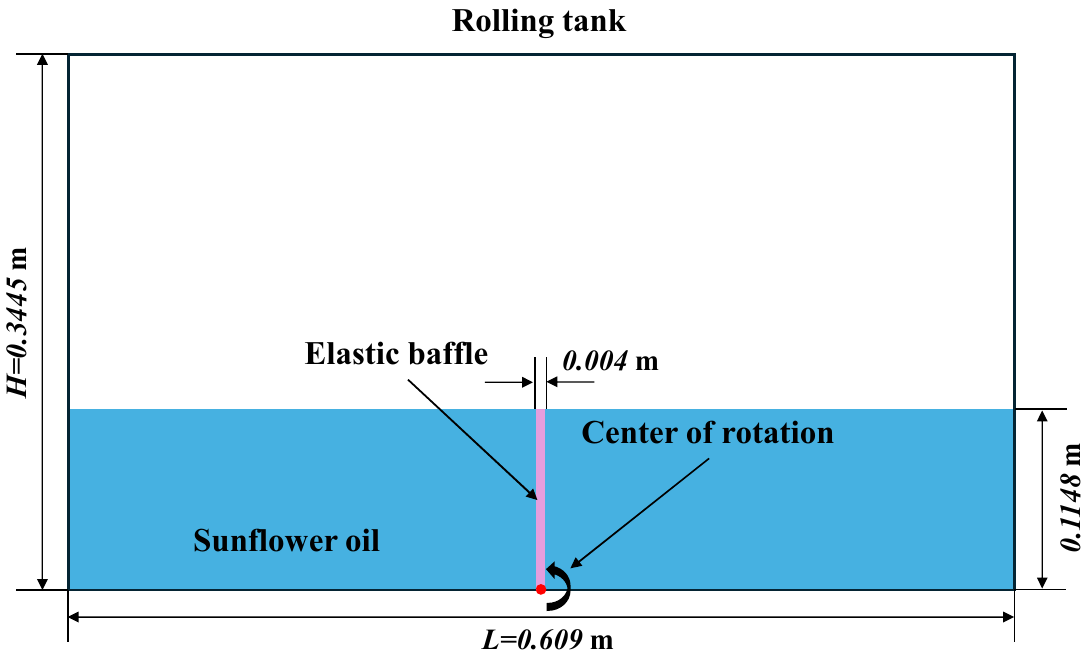}
	\caption{Sloshing in a rolling tank with a elastic 
		baffle: Schematic illustration.}
	\label{sloshingbaffleconfiguration}	
\end{figure}

Finally, the proposed RKGC method is validated using a rolling 
tank example, where an elastic baffle is clamped at the bottom.
Following the experimental setup performed by Idelsohn et al.
\cite{idelsohn2008interaction}, the computational configuration 
is illustrated in Fig.~\ref{sloshingbaffleconfiguration}.
The tank, measuring $0.609\times0.3445$~m, is filled with 
sunflower oil to a height of $0.1148$~m. 
The density and kinematic viscosity of the oil are $\rho^F=917\ 
\mathrm{kg/m^3}$ and $\mu^F=5.0\times10^{-5}\ \mathrm{m^2/s}$, 
respectively. 
An elastic baffle made of electrolyte polyurethane is clamped 
at the bottom center of the tank, with its height matching 
the liquid level. 
The baffle has a thickness of $b=0.004$~m and a density of 
$\rho^S =1100\ \mathrm{kg/m^3}$. 
Its mechanical properties include Young's modulus of 
$E^S=6.0\mathrm{MPa}$ and a Poisson’s ratio of $\nu^S=0.49$.
The tank undergoes rolling motion about its bottom center 
with a maximum amplitude of $4^{\circ}$ and a period of 1.211~s.
A sensor at the free end of the baffle tracks the horizontal 
displacement. 
Simulations use the particle spacing ratio of $b/dp^S=4$.

\begin{figure}[htb!]
	\centering
	\includegraphics[width=1\textwidth]
	{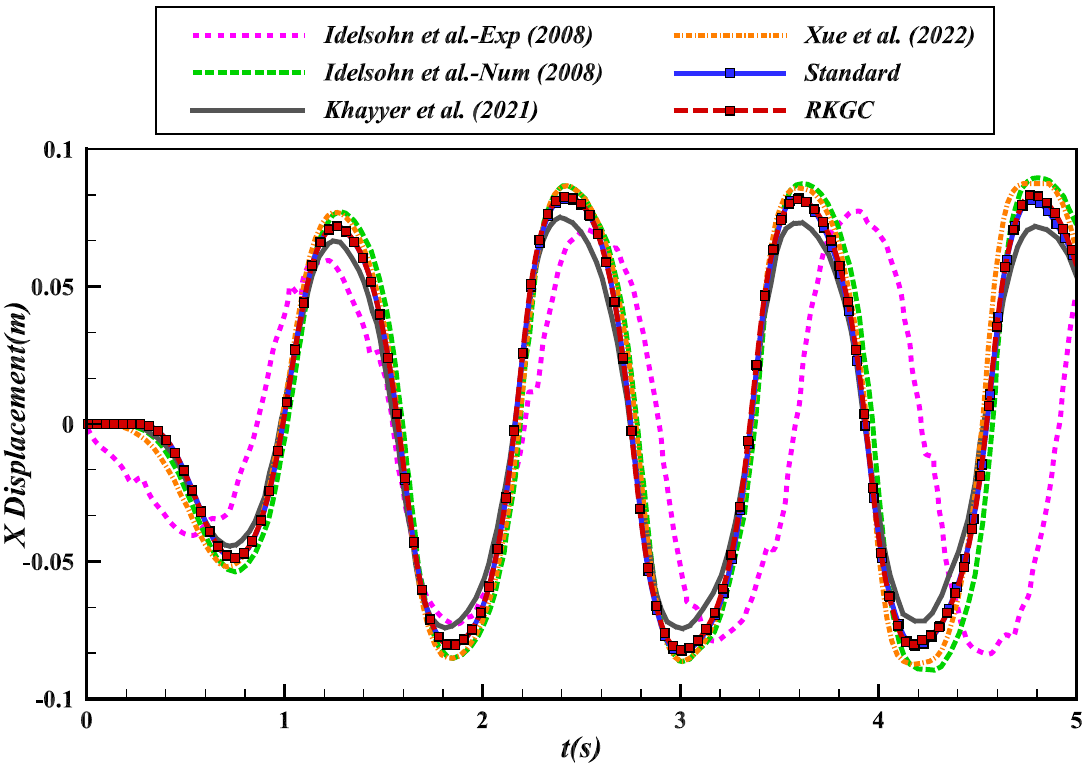}
	\caption{Sloshing in a rolling tank with a elastic 
		baffle: Time history of free end horizontal 
		displacement of the elastic baffle .}
	\label{sloshingelasticbaffledisplacement}	
\end{figure}

Fig.~\ref{sloshingelasticbaffledisplacement} compares time 
history of the baffle’s horizontal displacement from the 
present method with experimental data~\cite{idelsohn2008interaction} 
and numerical results from Khayyer et al.~\cite{khayyer2021multi} 
and Xu et al.~\cite{xue2022novel}. 
After an initial transient phase, the baffle exhibits regular 
oscillations with stable amplitude and period. 
Both standard and RKGC SPH methods align with reference results 
in amplitude and frequency, validating the good accuracy of the 
RKGC method.
However, observed discrepancies in swing period frequency, also 
noted in Refs.~\cite{idelsohn2008interaction, xue2022novel}, 
may stem from the inertial effects.

\begin{figure}[htbp]
	\vspace{-3cm}
	\centering
	\includegraphics[width=0.92\textwidth]
	{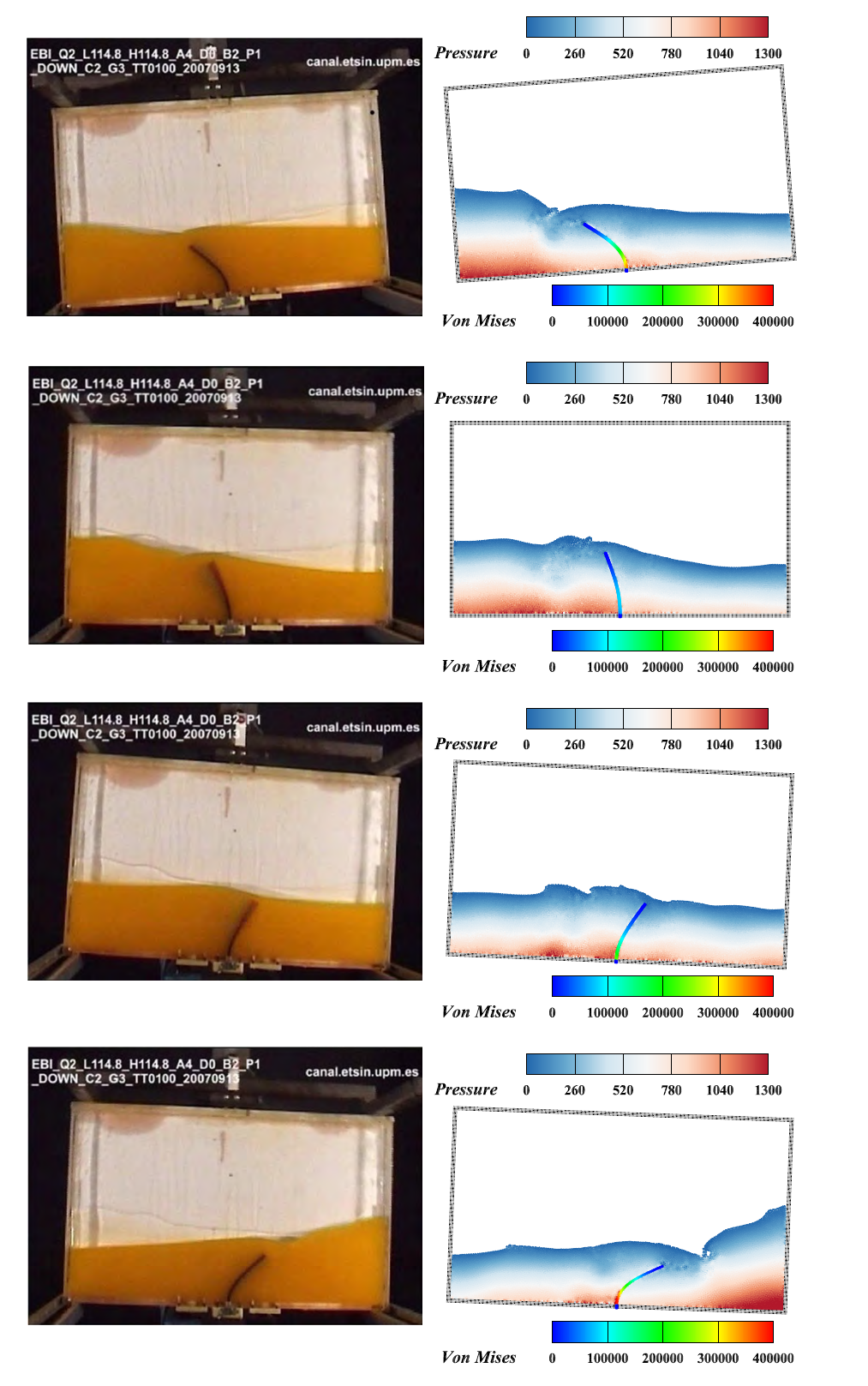}
	\caption{Sloshing in a rolling tank with a elastic 
		baffle: Comparison 
		between simulation results obtained by the RKGC method 
		and experimental results~\cite{idelsohn2008interaction} 
		at $t=1.84$~s, $2.12$~s, $2.32$~s and $2.56$~s.}
	\label{sloshingelasticbafflecontour}	
\end{figure}

Fig.~\ref{sloshingelasticbafflecontour} presents snapshots of 
the sloshing behavior in the rolling tank with the elastic 
baffle obtained using the RKGC method. 
These are compared with experimental observations from 
Ref.~\cite{idelsohn2008interaction} at $t=1.84$~s, $2.12$~s, 
$2.32$~s and $2.56$~s.
The elastic baffle effectively isolates the pressure field 
within the liquid, serving as a dynamic motion boundary. 
Its presence also disrupts the free surface, leading to an 
irregular 
morphology. 
The comparison of numerical and experimental results at 
various time intervals show strong agreement in both the free 
surface shape and structural deformation, further validating 
the accuracy of the proposed numerical model.
%
%
%
\section{Conclusion and remark}\label{conclusion}
From the perspective of fluid-structure interaction (FSI) 
challenges, this study introduces a corrected Riemann SPH 
method for the fluid domain to enhance numerical accuracy 
in multi-resolution FSI simulations.
Integrated with the RKGC formulation, which ensures second-order 
convergence and first-order consistency in SPH approximations, 
the fluid domain is solved with the corrected Riemann SPH method, 
which plays a significant role in multi-resolution FSI analysis, 
where the fluid domain generally has low resolutions.
Validation across various FSI cases confirms that the proposed 
RKGC-corrected Riemann SPH method improves accuracy and ensures 
robust convergence.
Future efforts will focus on refining fluid-solid interaction 
treatments and extending high-order consistency corrections to 
the solid phase.
Addressing these aspects will further establish SPH as a 
versatile, efficient, and high-fidelity tool for modeling 
complex FSI phenomena in engineering applications.
%
%
%
\section*{Acknowledgments}
\addcontentsline{toc}{section}{Acknowledgment}
Bo Zhang acknowledges the financial support provided by 
the China Scholarship Council (No. 202006230071). 
X.Y. Hu expresses gratitude to the Deutsche 
Forschungsgemeinschaft (DFG) for sponsoring this research 
under grant number DFG HU1527/12-4. 
The corresponding code for this work is available on 
GitHub at \url{https://github.com/Xiangyu-Hu/SPHinXsys}.
\section*{Author Declarations}
The authors have no conflicts to disclose.
%
%
\bibliographystyle{elsarticle-num}
\bibliography{fsicorrection}
\end{document}